\documentclass[twocolumn,english,showpacs,showkeys,superscriptaddress,citeautoscript]{revtex4}
\usepackage[T1]{fontenc}
\usepackage[latin9]{inputenc}
\usepackage{color}
\usepackage{float}
\usepackage{amsmath}
\usepackage{amssymb}
\usepackage{graphicx}
\usepackage{esint}

\usepackage{ulem}

\makeatletter

\@ifundefined{textcolor}{}
{%
 \definecolor{BLACK}{gray}{0}
 \definecolor{WHITE}{gray}{1}
 \definecolor{RED}{rgb}{1,0,0}
 \definecolor{GREEN}{rgb}{0,1,0}
 \definecolor{BLUE}{rgb}{0,0,1}
 \definecolor{CYAN}{cmyk}{1,0,0,0}
 \definecolor{MAGENTA}{cmyk}{0,1,0,0}
 \definecolor{YELLOW}{cmyk}{0,0,1,0}
}

\usepackage{babel}

\usepackage{caption}
\usepackage{subcaption}

\makeatother

\usepackage{babel}

\begin{document}
\title{Anomalous thermodynamics in a mixed spin-1/2 and spin-1
hexagonal nanowire system}
\author{R. A. Pimenta}
\email{pimenta@ifsc.usp.br}
\affiliation{Departamento de F\'isica, 
          Universidade Federal de Lavras, 
          Caixa Postal 3037, 37200-000, Lavras, MG, Brazil}

\author{O. Rojas}
\email{ors@ufla.br}
\affiliation{Departamento de F\'isica, 
          Universidade Federal de Lavras, 
          Caixa Postal 3037, 37200-000, Lavras, MG, Brazil}

\author{S. M. de Souza}
\email{sergiomartinsde@ufla.br}
\affiliation{Departamento de F\'isica, 
          Universidade Federal de Lavras, 
          Caixa Postal 3037, 37200-000, Lavras, MG, Brazil}

\date{\today{}}

\begin{abstract}
The mixed spin-1/2 Ising model and spin-1 Blume-Capel model in an
hexagonal nanowire structure under the presence of crystal field
is considered. The free energy is obtained through the transfer matrix
technique, which is solved numerically. Our main result lies in the
presence of pseudo-transition in low temperature region near the ferrimagnetic/ferromagnetic
boundary, due to the influence of a crystal field. The evidence of a pseudo-transition is observed in several
quantities. Free energy first derivative quantities like entropy and
internal energy show an abrupt but continuous jump,
whereas quantities associated with second derivatives of the free energy like the
specific heat exhibit a strong sharp peak,
quite similar to a second order phase transition. We also investigate
magnetization patterns and do not find evidence of spontaneous magnetization.
Nevertheless, assuming a small magnetic field, we can induce a magnetization
which resembles a spontaneous magnetization at a pseudo-critical
temperature.
\end{abstract}
\maketitle

\section{Introduction}

Recently, the magnetic properties of cylindrical
nanowires have shown great potential applications such as magnetic recording,
shift registers and logic gates, among other possible candidates.
Indeed, over the last decades, several researches have been devoted
to examine the magnetic and thermodynamic properties of nanowire systems,
both experimentally \cite{Bran,Iorio,kmoon,ghaddar,torres,Wojcik}
and theoretically \cite{Kantar14,Mendes,Osman,keskin,Deviren,hachem,kocakaplan,kantar15,kantar-jmmm15,nmaila,holanda,mendes20}.

From the theoretical point of view, several techniques have been used,
such as mean-field approximation (MFA) \cite{Mendes,mendes20}, effective field
theory (EFT) \cite{Kantar14,kocakaplan,kantar15}, Monte Carlo simulation
(MC) \cite{mendes20,iglesias,vasilakaki}, among others \cite{Wesse,tanriv}.
Special attention has been given to the study of mixed spin systems, in particular those
composed by spin-1 and
spin-1/2 particles arranged in hexagonal structures \cite{Kantar14,Mendes,kantar15,kantar-jmmm15,mendes20}.
These works have predicted the occurrence of first and second order phase transitions
as well as spontaneous magnetization at finite temperature.
Such findings are astonishing, since these models do not violate the conditions
of the non-existence theorem for phase transitions in
one-dimensional systems with short range interactions, see {\it e.g.} \cite{cuesta}
and references therein. Here the non-existence of phase transitions follows from
the Perron-Frobenius
theorem \cite{Ninio,ky lin} when applied to the associated transfer matrix \cite{Baxter}.
Therefore, it is clear that one must be careful in drawing conclusions from approximative
methods for such systems. Very recently, for example, the 
the failure of EFT to predict phase transitions in zero-dimensional and
one-dimensional models has been thoroughly discussed
in reference \cite{strecka21}.  

Nevertheless, peculiar anomalous behavior in thermodynamic quantities
which resembles phase transition has been observed in a variety of
1D models, such as the Ising-Heisenberg spin model
on a triangular tube \cite{strk-cav}, the diamond chain structure \cite{psd-Ising,unv-cr-exp,rojas21,Krokhmalskii},
the tetrahedral chain \cite{thetra-hedr,Galisova} and even in a genuine
one-dimensional Potts model \cite{panov}. The anomalous behavior
has been called a pseudo-transition \cite{souzarojas18} and it is
characterized by a continuous steep change (quite
similar to a discontinuity) in the first derivatives of the free
energy and by a giant peak (very similar to a singularity)
in its second derivatives. Even so, according to the Perron-Frobenius theorem \cite{Ninio,ky lin}, these one-dimensional
models do not contradict the non-existence phase transition theorem
since
the transfer matrix elements are strictly positive.

\begin{figure}
\includegraphics[scale=0.6]{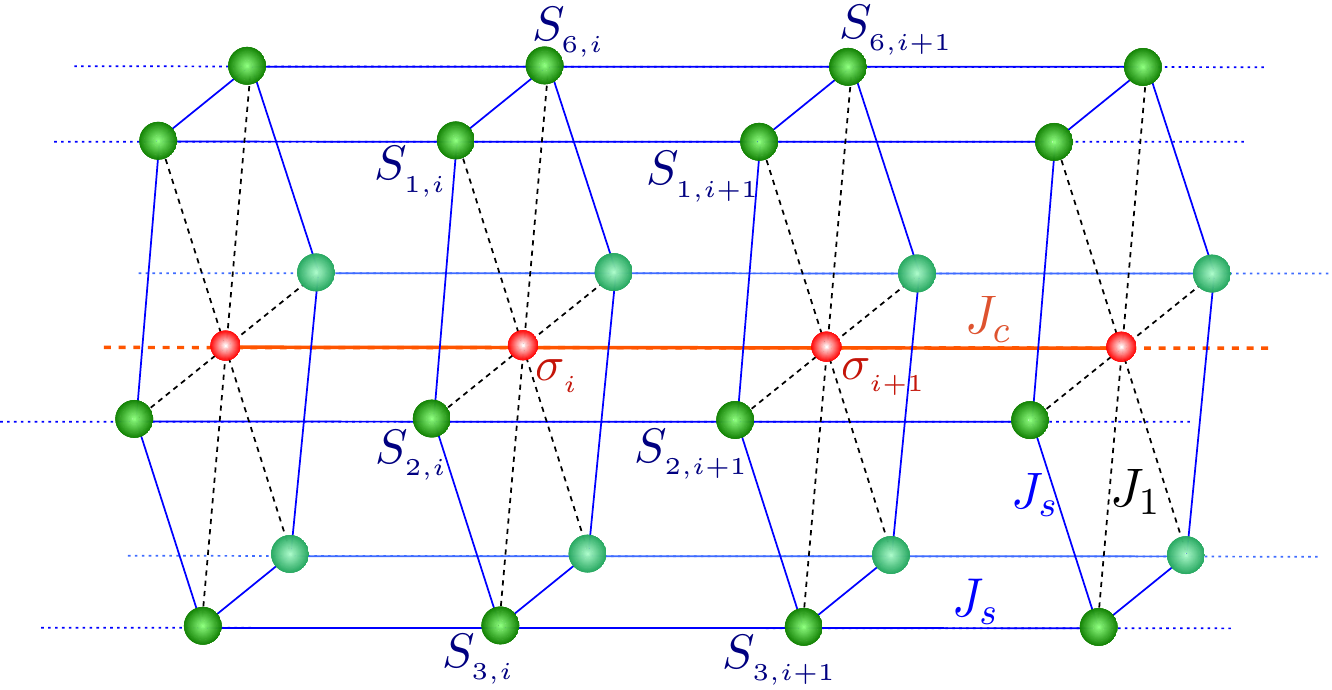}\caption{\label{fig:nanowire}Mixed magnetic nanowire with spin 1/2 Ising in
the central column and spin 1 Blume-Capel in the external hexagonal
layer.}
\end{figure}

Within in this context, we were motivated to analyze a 1D model
in a hexagonal structure using the transfer matrix technique, which provides exact numerical
results. Precisely, we investigate rigorously a mixed
hexagonal Ising type nanowire system, composed by spin-1/2 particles
in the core and spin-1 particles in the shell, previously
considered in references \cite{Kantar14,Mendes,mendes20}, see fig.\ref{fig:nanowire}.
For this nanowire system it has been conjectured the existence of a possible
first and second order type phase transitions, as well as possible
spontaneous magnetization for certain values of the model parameters. As expected,
the results from the transfer matrix approach do not confirm this conjecture. However,
we do find a quite anomalous thermodynamic behavior.

Let us recall that the powerfulness of the
transfer matrix approach for 1D systems relies in
the fact that it allows to extract physical information in the thermodynamic limit
from a finite sized matrix. However, for the model depicted
in fig.\ref{fig:nanowire} the transfer matrix is huge, with dimension $1458\times 1458$.
This is probably the reason why 
this approach has not yet been used, to the best of our knowledge,
to study 1D nanowire systems. In fact, closed solutions are out of range
and even numerical studies are computationally demanding. Even
with good computing resources but using standard numerical programming
languages such as Fortran, C++, Python, and other numerical computing
software, they are usually limited to 15-18 decimal digits of precision,
making the transfer matrix technique non-attractive. Moreover, especially
in low-temperature regions, performing a high-precision numerical
computation is hardly feasible using these languages. Consequently, obtaining physical quantities
involving free energy derivatives becomes challenging because a naive
analysis might induce us to a spurious outcome. In this sense, to
get the maximum power of the numerical transfer matrix approach, we
need a high precision numerical computation far beyond the 15-18 decimal
significant precision. Hence, we use computer algebra systems like {\tt Mathematica} and
{\tt Maple},
which provides arbitrary digits precision according
to demand; typically, we use up to 50 digits or more. Indeed, this
result is enough to perform any numerical derivatives of free energy
with confidence avoiding misleading conclusions. 

The use of high precision numerics
allowed us a careful analysis of the anomalous behavior in various thermodynamic
quantities in the nanowire model. For example, we do observe abrupt changes in the internal
energy, entropy and magnetic quadrupole. On the other hand, we observe
sharp peaks in the specific heat and correlation length. 
Indeed, a naive analysis could induce us to affirm a truly phase transition.
However, once the thermodynamics is analyzed with magnifying glass, one concludes
that this anomalous behavior is nothing else than a pseudo-transition
observed previously in \cite{souzarojas18}. The pseudo-transition
occurs near a ferrimagnetic (or core-ferromagnetic)
and ferromagnetic quasi-phase boundary.

Let us remark that while the pseudo-transition can be visually observed in several thermodynamic quantities,
the precise determination of the quasi-phase boundary is a delicate issue since there
is no truly phase transition. For example, in a true phase transition, there is a divergence in the specific heat and correlation length at a well defined critical temperature. In the present model, there is no singularity in these quantities, although the peaks are sharp and
high. Within this limitation, we determine the temperatures where 
the peaks in the third largest eigenvalue of the transfer matrix (linked to the correlation function), $T_p$, and the specific heat, $T_{p'}$, occur, and also the temperature
where a simple and interesting relation is satisfied by the eigenvalues of the transfer matrix,
$T_0$. We propose that the quasi-phase boundary is determined by the condition
$T_{0}\approx T_{p}\approx T_{p'}$.

This paper is organized as follows. In Section \ref{sec:ham}, we
define the nanowire Hamiltonian and analyze its ground state phase
diagram. Next, in Section \ref{sec:transfer}, we introduce the transfer
matrix, and carefully analyze its eigenvalues behavior.
In Section \ref{sec:thermo}
we present the various thermodynamic quantities obtained from the
numerical diagonalization of the transfer matrix, and discuss the quasi-phases of the system. In
\ref{sec:conclu} we present the conclusions. In the appendix \ref{app}, we make some considerations
about the numeric derivatives.

\section{Nanowire Hamiltonian}\label{sec:ham}

The nanowire model we consider is a mixed spin system
composed of an Ising spin 1/2 (red circles) coupled to  
Blume-Capel spin-1 (green circles) per unit cell, see fig.\ref{fig:nanowire}. The system can be
described by the following Hamiltonian, 
\begin{equation}
H=\sum_{i=1}^{N}H_{i,i+1},
\end{equation}
with
\begin{alignat}{1}
H_{i,i+1}= & -\sum\limits _{j=1}^{6}\left[J_{1}S_{j,i}\sigma_{i}+J_{s}\left(S_{j,i}S_{j,i+1}+S_{j,i}S_{j+1,i}\right)\right]\nonumber \\
 & -J_{c}\sigma_{i}\sigma_{i+1}-\sum\limits _{j=1}^{6}\left(DS_{j,i}^{2}+h_{s}S_{j,i}\right)-h_{c}\sigma_{i},\label{eq:H}
\end{alignat}
where $\sigma_{i}\in\left\{ -\frac{1}{2},\frac{1}{2}\right\}$ and $S_{j,i}\in\left\{ -1,0,1\right\}$ are the spins at the lattice
site $i$ with $j$ marking the position in the i-th hexagon, and we
assume periodic boundary conditions.
The coupling constants are 
 $J_{1}$ between Ising and Blume-Capel spins,
$J_{c}$ between Ising spins and $J_{s}$ between Blume-Capel spins, $D$ denotes
the single-ion anisotropy or crystal field parameter of Blume-Capel spins, and $h_c$ and $h_s$ are external magnetic fields. 

\subsection{Zero temperature phase diagram}

In this section, we examine the phase diagram of the
Hamiltonian (\ref{eq:H}) at zero temperature. Since there
are many free parameters in (\ref{eq:H}), we expect
that the phase diagram of model has a rich structure. Motivated
by results obtained using the mean field approximation
\citep{Mendes},
we analyze the model around the crystal field value $D=-2.5$. 
In this work,
for simplicity, we fix part of the set of parameters
to be $J_{1}=J_{s}=J_{c}=1$, according to \citep{Mendes}. 

\begin{figure}[h!]
    \centering
  \includegraphics[width=8cm]{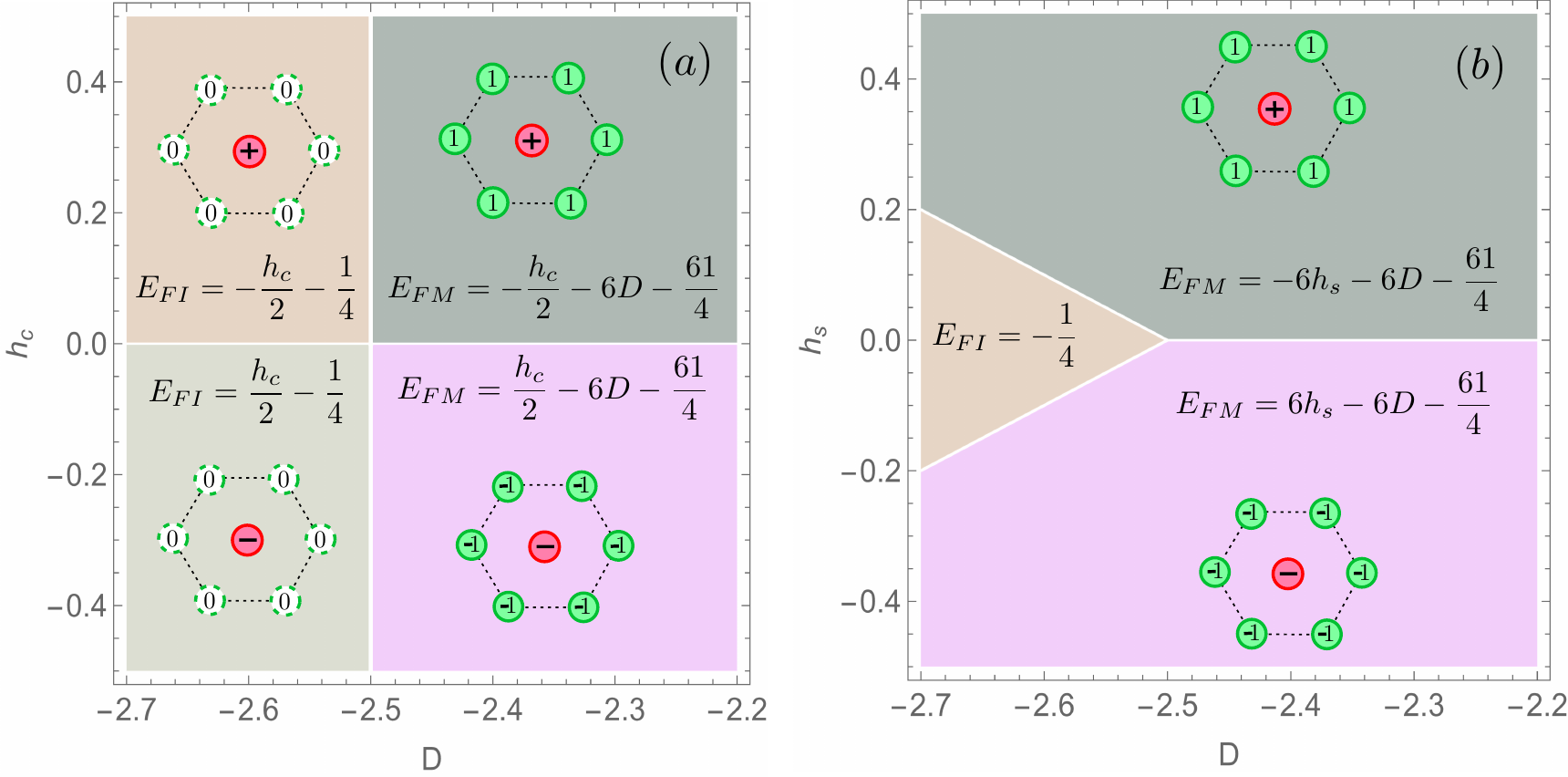}
    \caption{\label{fig:ground}Ground state phase diagram at zero temperature. (a) $h_c \times D$ for $h_s=0$. (b) $h_s \times D$ for $h_c=0$.}
\end{figure}

In fact, the work \citep{Mendes} predicts the existence of
first-order phase transition at finite temperature around $D=-2.5$.
This foretold first-order
phase transition supposedly emerges as a consequence of the phase
transition occurring at zero temperature. Therefore, we focus our attention
on the vicinity of the zero-temperature phase transition. We consider
the cases $D<-2.5$, $D=-2.5$ and $D>-2.5$, with and without magnetic
fields. The phase diagrams, as function of $D$, $h_c$ and $h_s$, are depicted in fig.\ref{fig:ground}.

For $D<-2.5$ and null magnetic fields $h_c=h_s=0$, the corresponding ground state is aligned with all
null shell spins, whereas the core spins are all aligned upwards or
downwards, setting a ferromagnetic state. For simplicity, let us call
this state as a ferrimagnetic (or core-ferromagnetic) state ($|FI\rangle$),
\begin{equation}
|FI\rangle=\prod_{i=1}^{N}|0,0,0,0,0,0\rangle_{i}\otimes|\sigma\rangle_{i},
\end{equation}
with corresponding ground state energy 
\begin{equation}
E_{FI}=-\frac{1}{4},
\end{equation}
so that the system will be doubly degenerate with $\sigma=\pm1/2$.  Since the system is strictly in the null magnetic fields,
we have the corresponding shell-spin magnetizations per unit cell $m_{s}=-\frac{1}{6}\tfrac{\partial E_{FI}}{\partial h_{s}}=0$ and core-spin
magnetization $m_{c}=-\tfrac{\partial E_{FI}}{\partial h_{c}}=0$.

For $D>-2.5$ and strictly null magnetic field $h_{c}=h_{s}=0$, the
system is in the ferromagnetic (FM) phase and ground state energy
is given by
\begin{alignat}{1}
E_{FM}= & -\frac{61}{4}-6D.
\end{alignat}
Similarly to our previous argument, the corresponding ground state
can be expressed as follows,
\begin{equation}
|FM\rangle=\prod_{i=1}^{N}|S,S,S,S,S,S\rangle_{i}\otimes|\sigma\rangle_{i},
\end{equation}
where $S=1$ and $\sigma=1/2$ or $S=-1$ and $\sigma=-1/2$. 
Both core and shell spins are fully parallel to each other, and each
of them doubly degenerates; therefore, the corresponding magnetizations become
$m_{s}=-\frac{1}{6}\tfrac{\partial E_{FM}}{\partial h_{s}}=0$ and $m_{c}=-\tfrac{\partial E_{FM}}{\partial h_{c}}=0$.

For $D=-2.5$ and null magnetic fields ($h_{c}=h_{s}=0$) the ground
state energy on the boundary is four-fold degenerate and its
is given by $E_{b}=-1/4$, with magnetizations
$m_{c}=m_{s}=0$.

We now analyze the effect of the magnetic fields $h_c$ and $h_s$, according
to fig.\ref{fig:ground}, panels (a) and (b) respectively.

For $D<-2.5$,
by turning on the external magnetic field ($h_{c}$), see fig.\ref{fig:ground}-(a),
the ground state energy will become,
\begin{equation}
E_{FI}=\begin{cases}
-\frac{1}{4}-\frac{1}{2}h_{c}, & h_{c}>0\\
-\frac{1}{4}+\frac{1}{2}h_{c}, & h_{c}<0
\end{cases},
\end{equation}
which implies that the system immediately raises its degeneracy by aligning
itself with the magnetic field $h_{c}$; thus, the core-spin magnetization
per unit cell leads to $m_{c}=-\tfrac{\partial E_{FI}}{\partial h_{c}}=1/2$ for
$h_{c}>0$ and $m_{c}=-1/2$ for $h_{c}<0$. Nevertheless, we must be careful when taking $h_{c}\rightarrow0$,
which apparently leads us to a different nonzero magnetization, contradicting
the magnetization with a strictly null magnetic field. Further details
on this issue will be discussed later when we examine the thermal
excitation.

For $D>-2.5$, fig.\ref{fig:ground}-(a), when we consider the external magnetic field $h_c\neq 0$, the ground
state energy is given by,
\begin{equation}
E_{FM}=
\begin{cases}
-\frac{61}{4}-6D-\frac{1}{2}h_{c}, & h_{c}>0\\
-\frac{61}{4}-6D+\frac{1}{2}h_{c}, & h_{c}<0\\
\end{cases},
\end{equation}
and it follows that the corresponding magnetization
will be $m_{c}=\pm1/2$. 

Now let us consider the effect of the field $h_s$,  fig.\ref{fig:ground}-(b).
Inside the triangle  $D+h_s\leq -2.5,D-h_s\leq -2.5 $ the system is a ferrimagnetic
state with ground state energy,
\begin{equation}
E_{FI}=
-\frac{1}{4}, \quad D+h_s\leq -2.5,D-h_s\leq -2.5 \,,
\end{equation}
and we have $m_c=m_s=0$. Outside the triangle, the degeneracy is lifted and
the ground state energy is given by,
\begin{equation}
E_{FM}=
\begin{cases}
-\frac{61}{4}-6D-6h_{s}, & h_{s}> 0\,,D+h_s>-2.5\\
-\frac{61}{4}-6D+6h_{s}, & h_{s}<0\,,D-h_s>-2.5\\
\end{cases},
\end{equation}
leading to the magnetization per unit cell
$m_s=\frac{1}{6}(\pm 6) = \pm 1$ and $m_c=0$.

Let us mention that one can turn on both fields $h_c$ and $h_s$ and find
interesting phase diagram. This point will be investigated elsewhere.

\section{Nanowire transfer matrix}\label{sec:transfer}

The transfer matrix technique is the main tool to find rigorous exact
results from the algebraic and numerical points of view. It has been
widely used since almost a century ago to obtain exact results mainly in
one and two two-dimensional lattices. Here we resort
to this technique to investigate the free energy of the nanowire model. Our principal
task is to solve the eigenvalues of transfer matrix $\mathbf{V}$
with elements,
\begin{equation}\label{transfer}
\langle\sigma,\{S_{i}\}|\mathbf{V}|\{S'_{i}\},\sigma'\rangle={\rm e}^{-H_{i,i+1}/k_{B}T},
\end{equation}
where $\{S_{i}\}$ denotes $\{S_{1},S_{2},S_{3},S_{4},S_{5},S_{6}\}$.
The full transfer matrix of the model given by (\ref{eq:H}) has dimension $1458\times1458$. 

In order to solve the transfer matrix more handily, we use the cylindric
symmetry $C_{6v}$ on the shell spins (composed by spin-1 particles).
The system is invariant under the rotation in $2\pi/6$ radians, so
we have
\begin{alignat}{1}
\mathbf{C}_{6}|S_{1},S_{2},S_{3},S_{4},S_{5},S_{6}\rangle= & |S_{2},S_{3},S_{4},S_{5},S_{6},S_{1}\rangle,
\end{alignat}
where the cyclic group $\mathbf{C}_{6}\mapsto\{I,\mathbf{C}_{6},\mathbf{C}_{6}^{2},\dots,\mathbf{C}_{6}^{5}\}$.
The system is also invariant under reflection plane containing the rotational nanowire axis,
so we have
\begin{alignat}{1}
\sigma_{v}|S_{1},S_{2},S_{3},S_{4},S_{5},S_{6}\rangle= & |S_{6},S_{5},S_{4},S_{3},S_{2},S_{1}\rangle,
\end{alignat}
where $\sigma_{v}\mapsto\{I,\sigma_{v}\}$.

Therefore, taking into account the $C_{6v}$ symmetry of the hexagonal
nanowire, we can express the spin-1 particles in the nanowire shell
by a matrix with dimension $729\times729$ which can be decomposed as a set of $12$ sub-block irreducible matrices, as described below
\begin{alignat}{1}
3^{6}= & 92\oplus89\oplus86\oplus3(\oplus80)\oplus6(\oplus37)\,.
\end{alignat}
By $n(\oplus v)$ we mean $\oplus v$ is repeated $n$ times. Including
the core spin, the transfer matrix dimension doubles, and the block
irreducible matrices dimension becomes as follows
\begin{equation}
3^{6}\otimes2=184\oplus178\oplus172\oplus3(\oplus160)\oplus6(\oplus74).
\end{equation}

As a result, we are able to express the transfer matrix of dimension
$1458\times1458$ into $12$ irreducible block matrices. We observe
that the largest eigenvalue of the transfer matrix comes from the
largest block matrix ($184\times184$). We can even use the spin inversion
symmetry in the absence of a magnetic field, so the largest matrix
may still shrink to $92\times92$. Here, however, we do not use
this symmetry since we need a non-null magnetic field
in order to analyze the magnetization.

Before considering the thermodynamic properties, let us check
some properties of the transfer
matrix eigenvalues. Let $\{\lambda_1,\lambda_2,\cdots,\lambda_{1458}\}$
denote the 1458 eigenvalues of the transfer matrix. In addition,
let us define for convenience
the normalized eigenvalues $\hat{\lambda}_{n}=\lambda_{n}{\rm e}^{-e_{0}/T}$,
where $e_{0}$ is the ground state energy.

\begin{figure}
\includegraphics[scale=0.6]{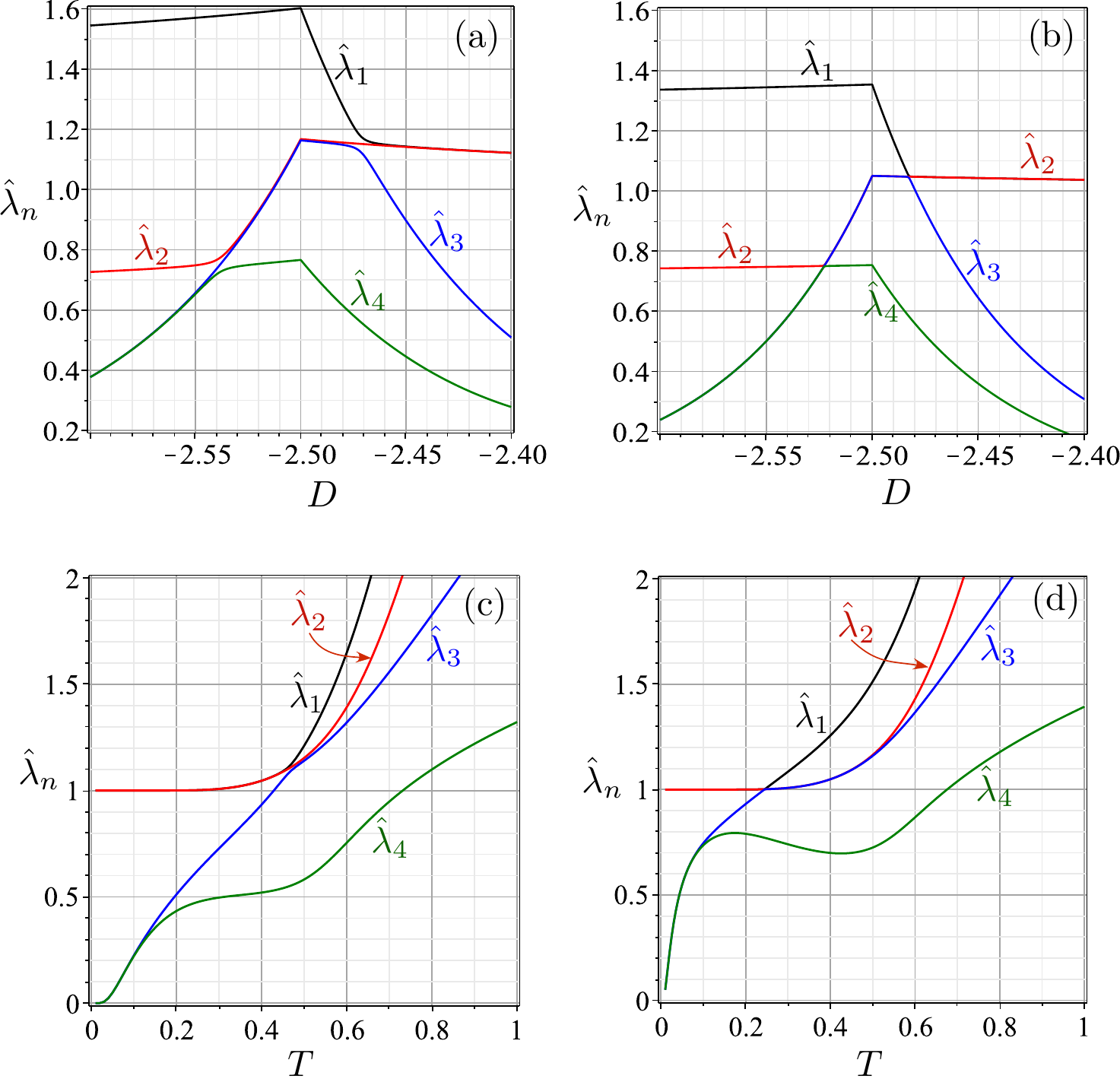} \caption{\label{fig:TM-eigs}Transfer matrix leading eigenvalues $\hat{\lambda}_{n}$ with $h_c=h_s=0$.
(a) For $\hat{\lambda}_{n}$ as a function of $D$, assuming fixed
$T=0.5$. (b) For $\hat{\lambda}_{n}$ as a function of $D$, assuming
fixed $T=0.4$. (c) For $\hat{\lambda}_{n}$ as a function of $T$,
assuming fixed $D=-2.495$. (c) For $\hat{\lambda}_{n}$ as a function
of $T$, assuming fixed $D=-2.475$.}
\end{figure}

In fig.\ref{fig:TM-eigs}
we illustrate the four largest eigenvalues around the field $D_{c}=-2.5$
where the zero temperature phase transition occurs. In panel (a) it is
depicted the four largest eigenvalues concidering a fixed temperature
$T=0.5$. We can observe that the curves $\hat{\lambda}_{1}$
and $\hat{\lambda}_{3}$ almost intersect for a given field $D_{p}$ in the interval
$-2.48<D<-2.47$. Also, we notice that 
for  $D<D_{p}$ the eigenvalues $\hat{\lambda}_{2}$ and $\hat{\lambda}_{3}$
are quasi degenerate, while for $D>D_{p}$ the eigenvalues $\hat{\lambda}_{1}$
and $\hat{\lambda}_{2}$ becomes quasi degenerate. At the same time,
we observe that $\hat{\lambda}_{4}$
is significantly smaller than the other three eigenvalues.

In fig.\ref{fig:TM-eigs}-(b), we plot the four largest normalized eigenvalues for the lower
temperature $T=0.4$. In this case, the curves do seem to intersect
at a given critical field $D_p$. Here, the curves $\hat{\lambda}_{1}$ and $\hat{\lambda}_{3}$ seems to form a continuous decreasing function which intersects with $\hat{\lambda}_{2}$. This feature seems to point out
to a phase transition at
this point where $\hat{\lambda}_{1}$, $\hat{\lambda}_{2}$ and $\hat{\lambda}_{3}$ becomes
almost the same for a particular value of parameter $D$.
Nevertheless, using high precision numerics, we can zoom in around this region
and observe actually a behavior similar to that depicted in panel (a), that is, all
curves are smooth.

We reserve panels (c) and (d) in fig.\ref{fig:TM-eigs} to analyze the behavior
of the four largest eigenvalues
as a function of the temperature for fixed values of $D$. In panel (c),
we fix $D=-2.475$, and observe that the curves $\hat{\lambda}_{1}$,
$\hat{\lambda}_{2}$ and $\hat{\lambda}_{3}$ tend to intersect
at a ``critical'' value $T_p\sim 0.48$.
Similarly, in panel (d), we consider the four largest eigenvalues
as a function of the temperature for fixed $D=-2.495$. That is, we get closer
to the zero temperature critical value $D_c=-2.5$. Here, virtually, the curves appear
to intersect. As before, the curves $\hat{\lambda}_{1}$
and $\hat{\lambda}_{3}$ appear to form a continuous function which is cut by  $\hat{\lambda}_{2}$. Nevertheless, the curves never cross, as high precision numerics
indicate.

Our high precision numerical results show that the largest eigenvalue
of the transfer matrix is non-degenerate, as expected the
Perron-Frobenius theorem \citep{Ninio,ky lin}. Further analysis of
transfer matrix eigenvalues will be given in the next subsection.

\subsection{Scaled eigenvalues behavior}

Recall that the matrix elements of the transfer matrix (\ref{transfer}) are
Boltzmann weights, that is, they are strictly positive for finite non-null temperature.
It follows from the
Perron-Frobenius theorem \citep{Ninio,ky lin} that the largest eigenvalue
of the transfer matrix is positive and non-degenerate. In addition to this
important fact, the deeper study of the transfer matrix eigenvalues provides
other crucial information about the physics of the system, see for instance
reference \citep{Lavis-2015}.

In fact, as we argue in the following,
certain properties of the transfer matrix eigenvalues
can be used to characterize
the boundaries between quasi-phases of the nanowire system.
Indeed, the behavior of the scaled eigenvalues, defined by,
\begin{equation}\label{scaledeig}
\Omega_n = \frac{\lambda_{n}}{\lambda_{1}}\,,
\end{equation}
is particularly insightful for $n=2,3$. Our numerical results
show that the behavior of the second and third largest eigenvalues
is relevant in the vicinity of $D_c=-2.5$. For example, one can notice that the
scaled eigenvalue have a strong peak
$\Omega_{3}\rightarrow1$ around $D_c=-2.5$, although never really attains $\Omega_{3}=1$. This
behavior is quite similar to that considered by Lavis \citep{Lavis-2015}
called an \textit{incipient phase transition}, defined for finite
size lattice. 

\begin{figure}[h]
\includegraphics[scale=0.6]{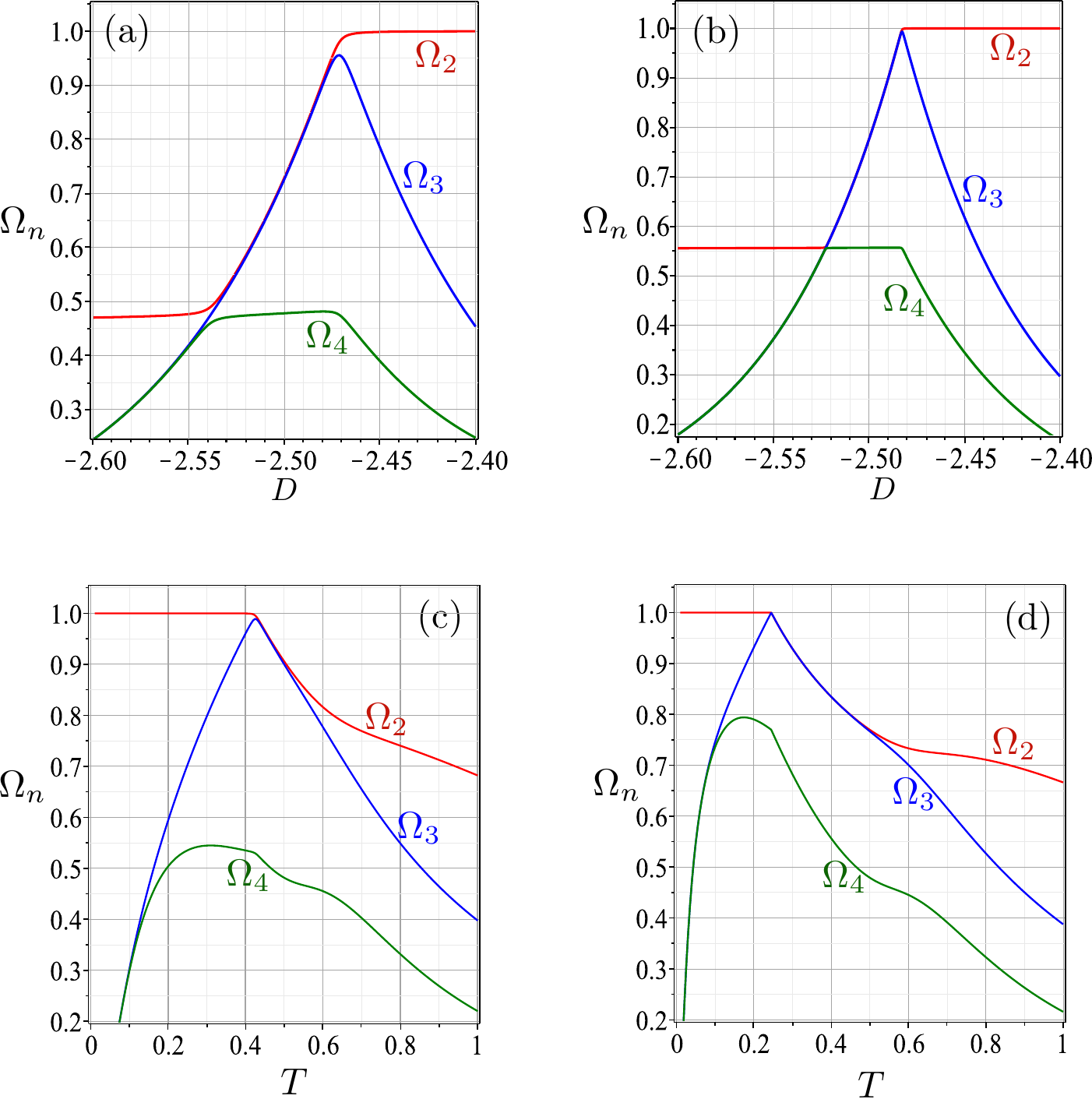}\caption{\label{fig:Largest-n}  (a) Largest scaled eigenvalues $\Omega_{n}$
as a function of $D$ with $h_c=h_s=0$ for $T=0.5$. (b) For $T=0.4$. (c) Same quantities
as a function of $T$ assuming
fixed $D=-2.48$. (d) For $D=-2.495$.}
\end{figure}

In fig.\ref{fig:Largest-n} we illustrate the scaled eigenvalues
$\Omega_{n}$ as a function of the parameter $D$, for fixed temperatures,
and as a function of $T$, for fixed values of $D$. The panel (a) reports
the temperature $T=0.5$, where we observe smooth curves, including
a
smooth peak in $\Omega_3$ around $D~-2.48$. For $D\gtrsim-2.48$
we observe that the largest and second largest eigenvalues are quasi degenerate,
\textit{i.e.}, $\Omega_{2}\rightarrow1$. The third largest eigenvalue
also plays an important role, that is, for a given parameter $D$
and a given finite temperature, $\Omega_{3}\rightarrow1$ exhibits
a maximum. Similarly, in panel (b), we consider the scaled eigenvalues
for a lower temperature $T=0.4$. These curves get sharper, indicating presumably
that there are crossing lines. However, the high-precision numerical
calculation shows smooth curves similar to that shown in panel (a).
These
curves still behave sharper for lower temperatures, and by using standard
numerical computation (like double-precision, providing 15 significant
digits), we cannot distinguish the eigenvalues $\lambda_{1}$ and
$\lambda_{2}$. Naively, it may be inferred that the two largest eigenvalues
are degenerate. However, high precision numerical computation
gives us far beyond than 16 significant digits; we can in fact easily handle
over 50 digits of precision using softwares like {\tt Mathematica} or {\tt Maple}, which clearly distinguishes the eigenvalues
$\lambda_{1}$ and $\lambda_{2}$, or $\lambda_{2}$ and $\lambda_{3}$,
as expected from the Perron-Frobenius theorem.
In panel (c), we report the same quantity $\Omega_{n}$ as a function
of temperature $T$ assuming fixed $D=-2.48$. Here we observe the quasi-degeneracy
between $\lambda_{1}$ and $\lambda_{2}$ roughly below the temperature of the peak of
$\Omega_{3}$. Indeed we can verify that all curves are smooth and
there is no degeneration between $\lambda_{1}$ and $\lambda_{2}$.
Likewise, in panel (d), we illustrate the eigenvalues for a fixed
value of $D=-2.495$, and the curves become sharper. A careless analysis
could induce the existence of crossing curves. However, a high-precision
numerical result confirms that crossing curves are not present.

\begin{figure}[h!]
    \centering
  \includegraphics[width=8cm]{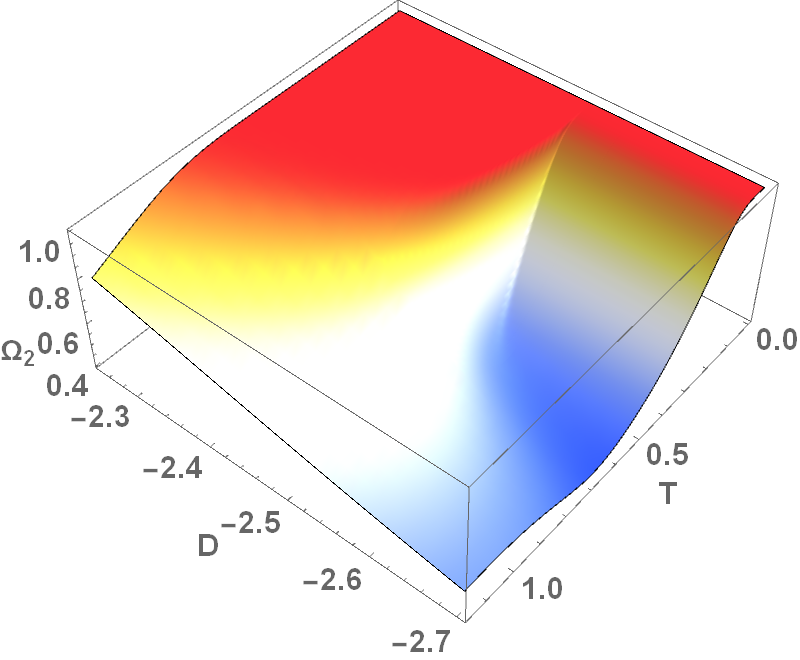}
    \includegraphics[width=8cm]{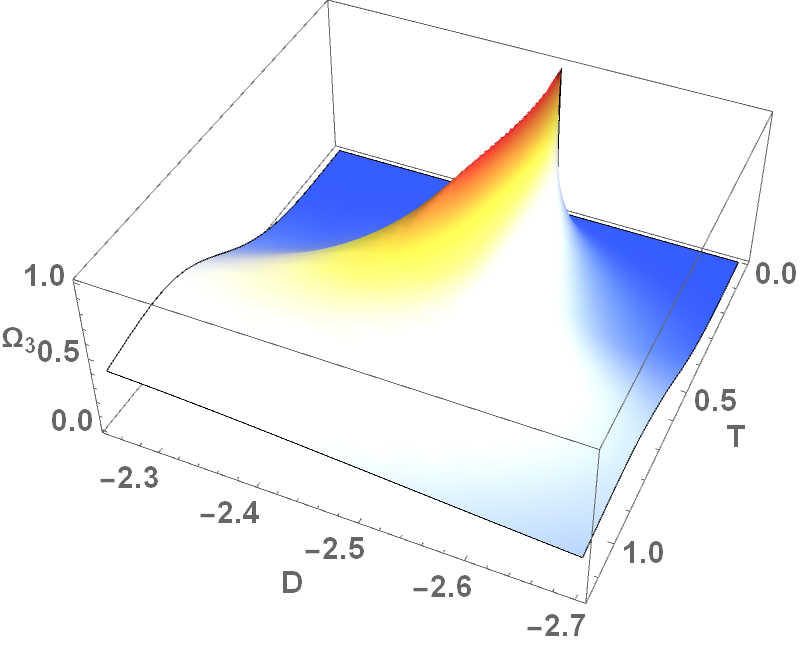}
    \caption{\label{fig:eig2D}The scaled eigenvalues $\Omega_2$ and $\Omega_3$ as a function of $D$ and $T$ for $h_c=h_s=0$.}
\end{figure}

It is also interesting to observe the behavior of the scaled eigenvalues
as a function of $D$ and $T$, as depicted in fig.\ref{fig:eig2D}.
For $\Omega_2$, we can observe a plateau $\Omega_2\sim 1$ for $D>-2.5$ which only
decays for relatively high temperatures. On the other hand, for $D<-2.5$ the decay occurs in the low temperature region. For $\Omega_3$, there is a very sharp region around $D_c=-2.5$ which,
interestingly enough, marks the separation between quasi-phases,
as we discuss below.

In summary, we cannot expect any phase transition at finite temperature,
neither first-order or second-order phase transition. However there
is an anomalous behavior induced by the single-ion anisotropy around $D_{c}=-2.5$, which we will discuss
below.

\section{Numerical thermodynamics}\label{sec:thermo}

In this section, we investigate numerically the dependence of the free energy
on the model parameters, as well as its first derivatives (entropy, internal energy, magnetization, magnetic quadrupole) and second derivatives (specific heat, susceptibility).  As before, for simplicity, we fix $J_{1}=J_{s}=J_{c}=1$. 

\subsection{Thermodynamic limit}

Let us recall that, for a finite size chain (finite $N$),
the partition function can be expressed
using all $1458$ eigenvalues of the transfer matrix, namely, 
\begin{equation}
\mathcal{Z}_{N}=\sum_{k=1}^{1458}\lambda_{k}^{N}\,.
\end{equation}
As a consequence, the free energy per unit cell can be expressed as follows 
\begin{alignat}{1}
f_{N}= & -T\ln\left(\lambda_{1}\right)-\frac{T}{N}\ln\left\{ 1+\sum_{n=2}^{1458}\Omega_n^{N}\right\} ,\label{eq:fN}
\end{alignat}
where $\Omega_{n}$ is defined by (\ref{scaledeig}) and
$T$ is the temperature in units of the Boltzmann constant.

All physical quantities will be analyzed in the thermodynamic limit. Since $\Omega_n<1$
for all $n>1$,
the free energy in thermodynamic limit ($N\rightarrow\infty$)
is given by,
\begin{equation}\label{fthermo}
f=-T\ln\left(\lambda_{1}\right).
\end{equation}
The equation (\ref{fthermo}) means that the largest eigenvalue of the transfer matrix
(which is a finite size matrix) captures the physics of the system in the thermodynamic limit.

Notice that although $\Omega_{2}$ and $\Omega_{3}$ are irrelevant within
the thermodynamic limit, these scaled eigenvalues are
important for the understanding of certain anomalies in the physical quantities.
For example, the peaks in the scaled eigenvalue $\Omega_{3}$ can be used
to characterize a pseudo-critical temperature transition.

\subsection{Free energy, entropy, internal energy and specific heat}

\begin{figure}[h!]
      \centering
 \includegraphics[width=8cm]{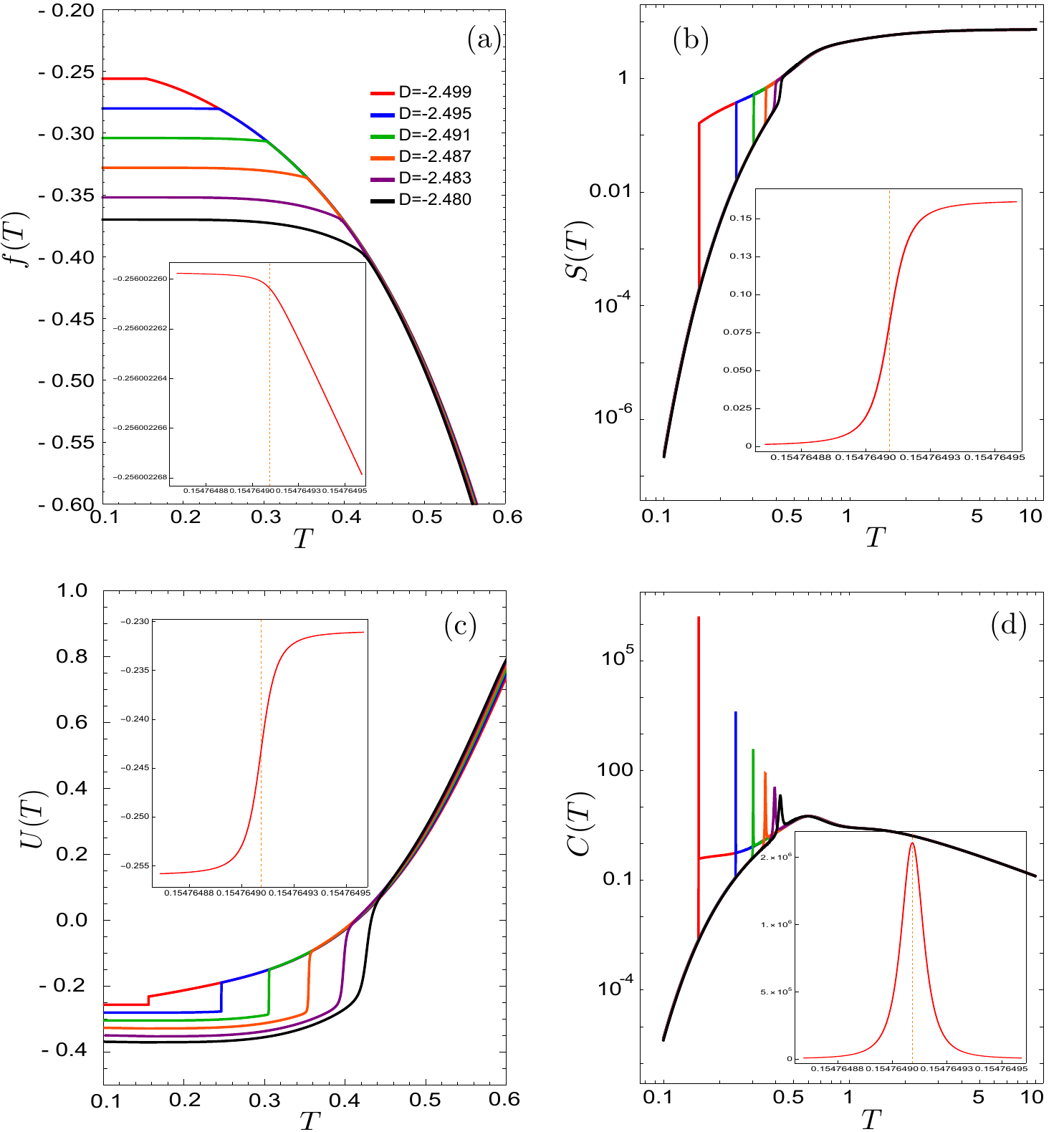}
        \caption{Thermodynamic quantities at zero fields $h_s=h_c=0$ for $0.1<T<10$. (a) Free energy. (b) Entropy. (c) Internal energy. (d) Specific heat. The free and internal energies are shown in a smaller range where they are visually distinguishable for different values of $D$.}
        \label{fig:geral}
\end{figure}

We start by considering the behavior of the free energy and its derivatives in the temperature range $0.1<T<10$ with zero fields $h_c=h_s=0$ and various single-ion fields $D$, namely, $D=-2.499,-2.495,-2.491,-2.487,-2.483,-2.480$. Numeric derivatives strategy is discussed in detail in appendix \ref{app}. The obtained results are shown in fig.\ref{fig:geral}. In panel (a), the free energy is depicted, and we can note that as one approaches $D_c=-2.5$, the free energy appears to have a ``corner'' around a given temperature. This ``corner'' smooths out as the field $D$ increases, and the bending of the curve occurs in higher temperatures. The apparent corner in the free energy is reflected in abrupt (but continuous) changes in the entropy, see panel (b), and internal energy, see panel (c),
as well as by the sharp and high peaks of
the specific heat, see panel (d). The discontinuity in the first and
divergence in the second derivatives, however, are in fact only apparent.
Indeed, by zooming in very close  the regions of abrupt changes, we do notice that all functions are continuous. As example, highlights for the field $D=-2.499$ are shown in fig.\ref{fig:geral}.

\begin{figure}[h!]
\centering
\includegraphics[width=8cm]{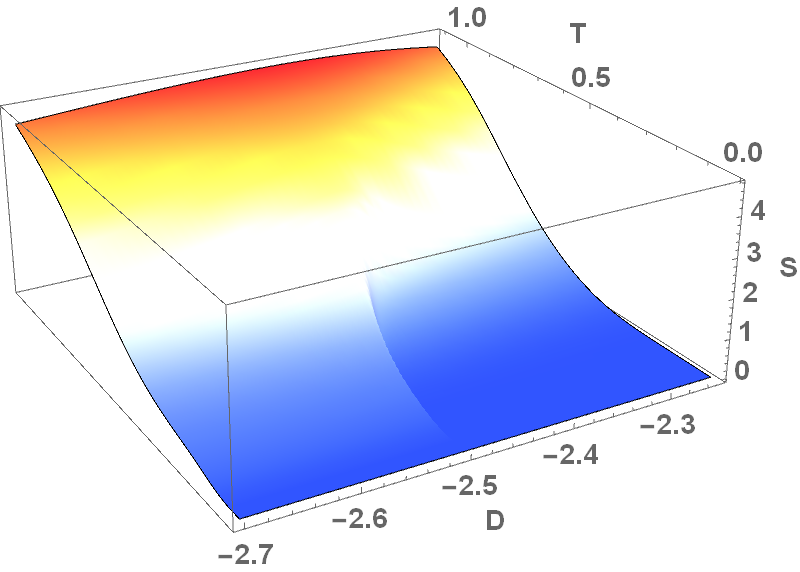}
\caption{\label{fig:entro2D}Entropy as a function of $D$ and $T$ for zero fields $h_c=h_s=0$.}
\end{figure}

It is worth looking at the entropy as a function of $D$ and $T$.
The entropy surface is shown in fig.\ref{fig:entro2D}. One can observe
that there is a apparent jump (discontinuity) in the entropy value around $D_c=-2.5$ separating the FI and FM phases. Since there is
no truly finite temperature phase transitions we define these regions by quasi-phases $qFI$ and $qFM$ \cite{unv-cr-exp,thetra-hedr,souzarojas18}. The quasi-phases reflect the
transition between the ferrimagnetic/ferromagnetic phases at zero-temperature.
Observe that, obviously,
there is no residual entropy
for both quasi-phases. But what about residual entropy at the boundary of
the two phases? We can verify that there is no residual entropy at
the interface because the system is only doubly degenerate, and there
is no macroscopic degeneration. When the residual entropy at the
phase boundary is null, it is possible to observe an anomalous
behavior at the phase boundary \citep{phs-bnd}, characterizing the pseudo-critical temperature.

\subsection{Correlation function}

As we discussed in Sec. III-a, the largest and the second largest eigenvalues are quasi degenerate in the low temperature region. Recall that the two point correlation function can be expressed in general as follows,
\begin{alignat}{1}
\langle O_{1}O_{r}\rangle= & \frac{\langle\psi|O_{1}V^{r-1}O_{r}|\psi\rangle}{\lambda_{1}^{r+1}}\nonumber\\
= & \langle O_{1}\rangle\langle O_{r}\rangle+\sum_{n=2}^{1458}\underbrace{\frac{\langle\psi|O_{1}|\psi\rangle\langle\psi|V^{r-1}O_{r}|\psi\rangle}{\lambda_{1}\lambda_{n}}}_{M_{n}}\Omega_n^{r}\nonumber\\
= & \langle O_{1}\rangle\langle O_{r}\rangle+\sum_{n=2}^{1458}M_{n}{\rm e}^{-r/\xi_{n}},
\end{alignat}
where the correlation length $\xi_{n}$ is given by 
\begin{equation}
\xi_{n}=-\frac{1}{\ln \Omega_n}.
\end{equation}

\begin{figure}[h!]
    \centering
   \includegraphics[width=8cm]{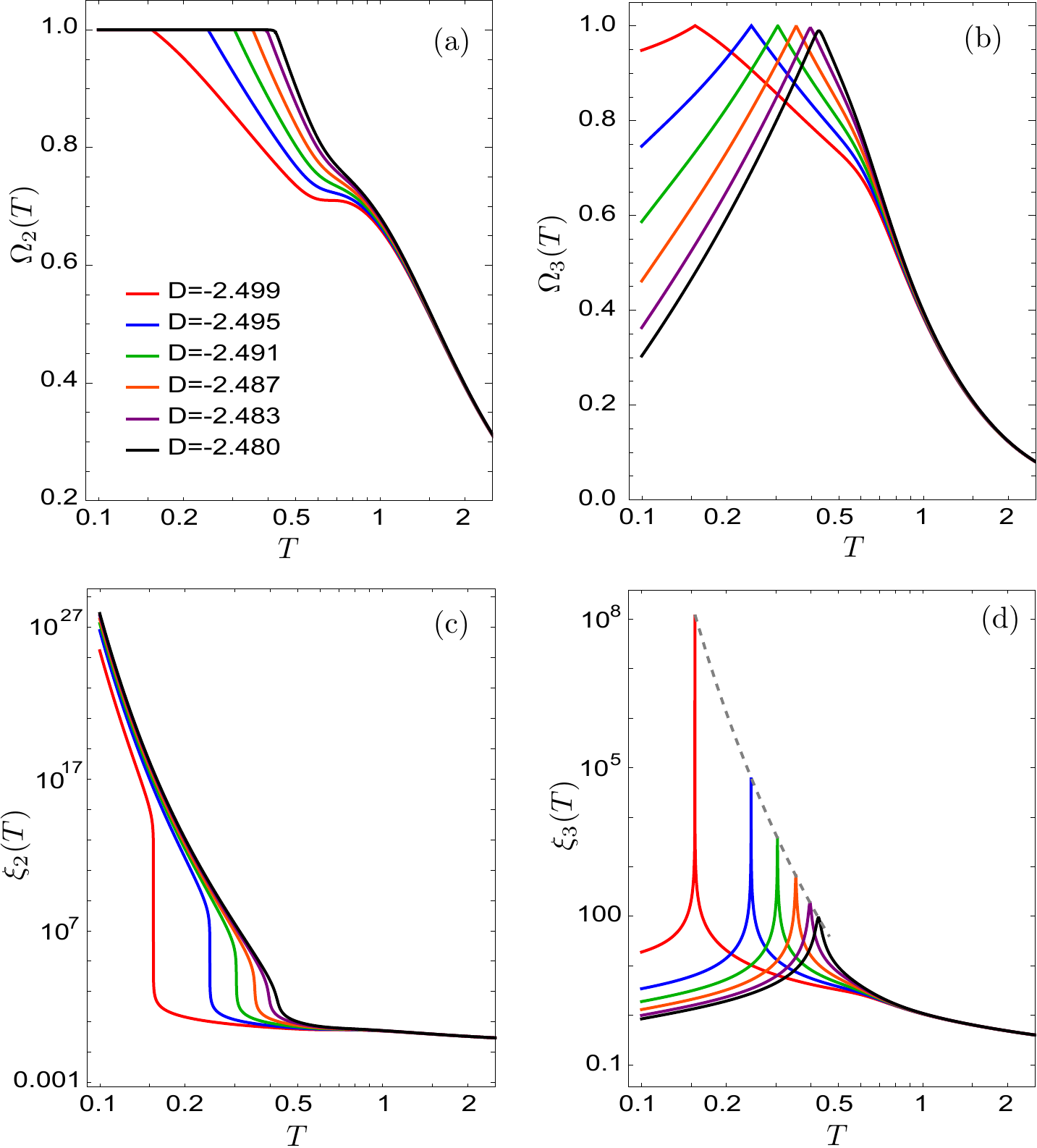}
        \caption{Ratios between the largest eigenvalues of the transfer matrix and associated correlation lengths for $h_c=h_s=0$. The dashed curve in gray is formed by the peaks of the correlation length for various values of $D$. (a) $\Omega_2(T)$. (b) $\Omega_3(T)$. (c) $\xi_2(T)$.
(d) $\xi_3(T)$.}
        \label{fig:eig}
\end{figure}

In fig.\ref{fig:eig}
we plot the scaled eigenvalues $\Omega_2$ and $\Omega_3$ as well as the correlations
$\xi_2$ and $\xi_3$ as functions of the temperature for different values of $D$. In panel (a), we confirm the quasi degeneracy of the largest and second largest
eigenvalues (that is, $\Omega_2\rightarrow 1$) below a given temperature. This is reflected in huge values of the
correlation $\xi_2(T)$, see panel (c). In panel (b), it is shown
the scaled eigenvalue $\Omega_3$ and we can observe the presence of well marked maximums $\Omega_3\rightarrow 1$. This is reflected in huge peaks of $\xi_3(T)$, see panel (d).
Despite the aggressive shapes, all curves are continuous. In fig.\ref{fig:eig}-(d), the curve
in dashed gray is formed by the correlation peaks obtained for various single-ion anisotropy
$D$.

\subsection{Quasi-phase and pseudo-critical temperature }

The results of the previous subsections indicate the absence of first
and second order phase transition for the nanowire model around $D_{c}=-2.5$,
as expected from the non-existence theorem for phase transitions in
one-dimensional systems with short range interactions \cite{cuesta}.
Nevertheless, there is a remarkable thermodynamic behavior, as depicted,
for example, in figs. \ref{fig:geral} and \ref{fig:eig}. These peaks
indicate the separation of two regions dominated by $FI$-type configuration
and $FM$-type configuration, following the zero temperature pattern
due to the thermal excitation in the low-temperature region.
The quasi-phase boundary is well marked for temperature $T\lesssim0.5$,
whereas this boundary smoothly vanishes for higher temperatures.

This behavior is characteristic of a pseudo-phase transition, and
it is interesting to determine the temperature where it occurs.
One possibility is that the temperature where the pseudo-phase transition occurs can be determined
from the peaks in correlation length $\xi_{3}$ or, equivalently,
from the peaks in the scaled eigenvalue $\Omega_{3}$.

Let us now analyze the peaks of scaled eigenvalue $\Omega_{3}$, making
the dependence on $T$ and $D$ explicitly. It is possible to find
a peak at a given temperature by setting $D$ and varying $T$. So,
it is enough to analyze the maximum of $\Omega_{3}(T,D)$ as follows:

\begin{equation}
\left(\frac{\partial\Omega_{3}(T,D)}{\partial T}\right)_{T_{p}}=\Omega_{3,T_{p}}=0,\label{eq:T_p}
\end{equation}

Alternatively, fixing $D$ we can find a temperature $T_{p'}$ where
occurs the peak of specific heat $C(T,D)$ at $T_{p'}$, that is 
\begin{equation}
\left(\frac{\partial C(T,D)}{\partial T}\right)_{T_{p'}}=C_{T_{p'}}=0.\label{eq:C_p}
\end{equation}
Here we perform precise numeric derivatives following  the information given in appendix \ref{app}.
\begin{figure}[h]
\includegraphics[scale=0.55]{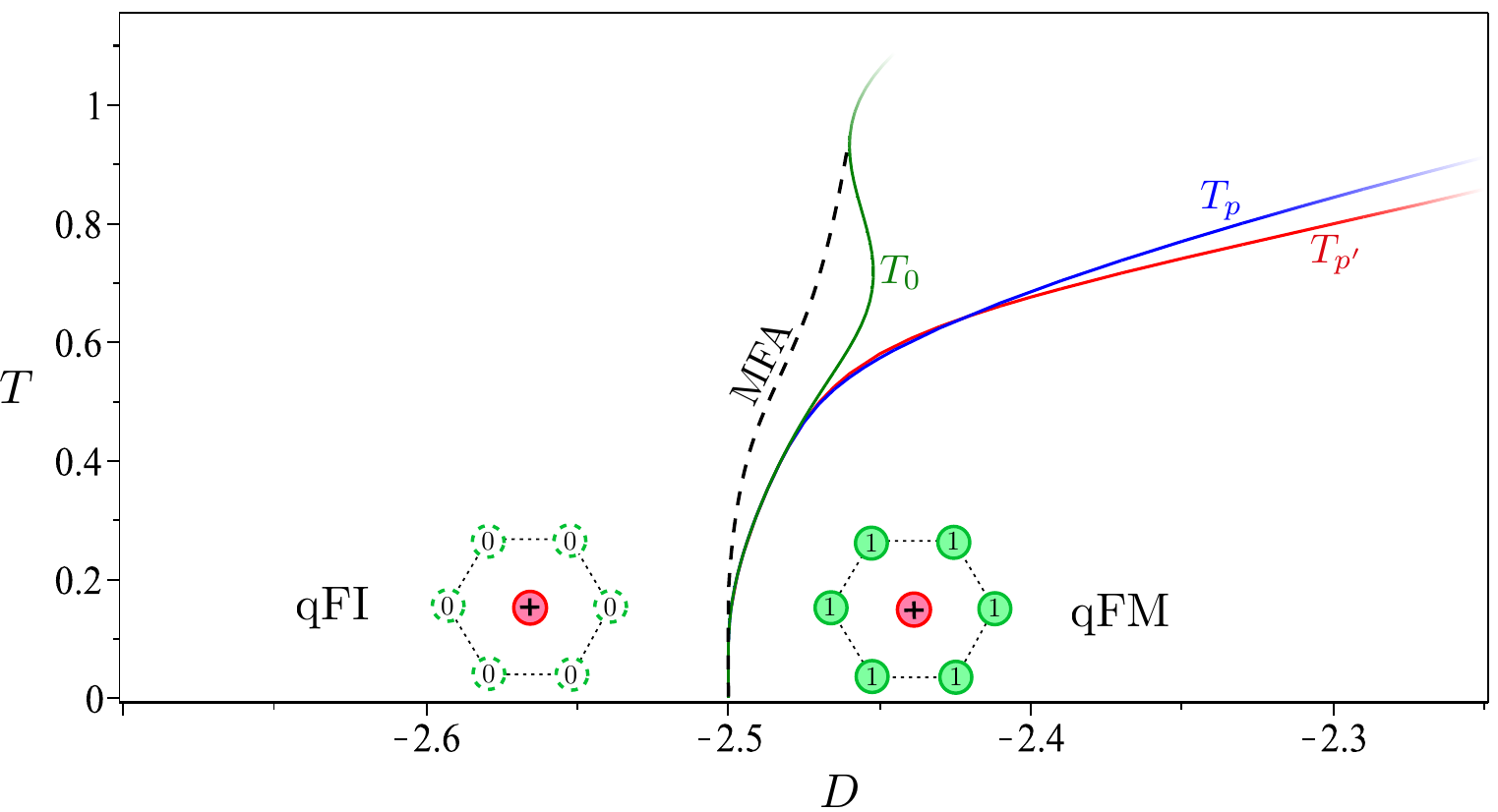}\caption{\label{fig:Pse-crit-temp}Pseudo-critical temperature $T_{p}$ as
a function of $D$, assuming zero magnetic field and fixed $J_{1}=J_{s}=J_{c}=1$,
using the different conditions defined in \eqref{eq:T_p} (blue curve),
\eqref{eq:C_p} (red curve) and \eqref{eq:L123} (green curve). The
dashed line describes the MFA result \citep{Mendes}.}
\end{figure}

Due to de absence of singularity, we cannot expect the sharp peaks
of correlation length and specific heat to satisfy the relation $T_{p}=T_{p'}$,
see fig.\ref{fig:Pse-crit-temp}. Although in the low-temperature
region, we observe $T_{p'}\rightarrow T_{p}$, particularly for the
parameters considered in fig.\ref{fig:Pse-crit-temp}, they are virtually
indistinguishable up to $T\approx0.5$. For higher temperatures, the
peak for both quantities occurs at different temperatures assuming
fixed $D$. For a more detailed comparison, we tabulated the highly
accurate $T_{p}$ and $T_{p'}$ in the second and third column of
Table \ref{tab:1}, respectively. To find highly accurate $T_{p'}$
reported in Table \ref{tab:1}, we need over 30 significant decimal
digits to perform precise numerical derivatives to obtain the specific
heat, then localize the temperature at the sharp peak. In table \ref{tab:1},
for $D\lesssim-2.499$, both quantities are equal up to order $10^{-15}$
(not shown), as we display only the first 13 decimal digits. Whereas,
for $D\gtrsim-2.495$, the peak temperatures differ in the tenth decimal
digit. As soon as $D$ increases, the differences become gradually
increasing.

\begin{table}
\begin{tabular}{|l|l|l|l|}
\hline 
$D$ & Eq.\eqref{eq:T_p}: $T_{p}$ & Eq.\eqref{eq:C_p}: $T_{p'}$ & Eq.\eqref{eq:L123}: $T_{0}$\tabularnewline
\hline 
\hline 
$-2.499$ & $0.1547649092435$ & $0.1547649092435$ & $0.1547649092435$\tabularnewline
\hline 
$-2.495$ & $0.2451364474144$ & $0.2451364476723$ & $0.2451364543420$\tabularnewline
\hline 
$-2.491$ & $0.3041836923345$ & $0.3041837182997$ & $0.3041850267111$\tabularnewline
\hline 
$-2.487$ & $0.3535672004682$ & $0.3535678022763$ & $0.3536037505689$\tabularnewline
\hline 
$-2.483$ & $0.3967956086305$ & $0.3968065691948$ & $0.3971636202687$\tabularnewline
\hline 
$-2.480$ & $0.4255221428649$ & $0.4255962726229$ & $0.4268513697425$\tabularnewline
\hline 
$-2.470$ & $0.4969631732231$ & $0.4994220556277$ & $0.5126192960524$\tabularnewline
\hline 
\end{tabular}\caption{\label{tab:1} Second column: peak temperature obtained from the ratio
of eigenvalues $T_{p}$. Third column: peak temperature of specific
heat. Fourth column: when eq.\eqref{eq:L123} is satisfied for a temperature
$T_{0}$, assuming $h_{c}=h_{s}=0$.}
\end{table}

Additionally, from eigenvalues behavior discussed in figs. \ref{fig:TM-eigs}
and \ref{fig:Largest-n}, we can establish the following condition
\begin{equation}
1+\Omega_{3}(T_{0},D)=2\Omega_{2}(T_{0},D),\label{eq:L123}
\end{equation}
such that fixing $D$ we can find a temperature $T_{0}$ that
satisfy \eqref{eq:L123}. The curve $T_{0}$ as a dependence of $D$
is illustrated in fig. \ref{fig:Pse-crit-temp} by a green line. Again,
we observe that $T_{0}\rightarrow T_{p}$ in low-temperature regions
is virtually identical up to $T\approx0.5$, while $T_{0}$ behaves
rather differently than $T_{p}$ for higher temperatures. Similarly,
in the fourth column of Table \ref{tab:1}, we observe in low-temperature
region $T_{0}$ leads to $T_{p}$ for $D\lesssim-2.499$, $T_{0}$ coincides
with $T_{p}$ up to 13 decimal digits, but when the crystal field $D$
increases, the difference becomes gradually evident.

As we can see, it is hard to characterize the anomalous peak, which
virtually looks like a genuine phase transition. In contrast, the
peak is sharply pronounced, but it gradually becomes a rounded, broad
peak as soon as $D$ increases. We cannot establish a limit between
sharp and broad peaks because the curves split slowly. A possible
definition of pseudo-transition could be the condition $T_{0}\approx T_{p}\approx T_{p'}$. In fig.\ref{fig:Pse-crit-temp}, this condition should occur
at $T_{p}\approx0.5$.

On the other hand, comparing with the ``critical''
temperature obtained from MFA \cite{Mendes,mendes20}, which is depicted
as a dashed line in fig.\ref{fig:Pse-crit-temp}, we observe that,
effectively, at sufficiently low temperature below $T\approx0.15$,
the mean-field result looks the same as $T_{p}$. Still, for higher temperatures,
the mean-field ``critical'' temperature
deviates considerably from the pseudo-critical temperature.

In summary, we do not observe either first or second-order phase
transition at a finite temperature, such as predicted in references \cite{Mendes,mendes20,Kantar14}.
Instead, we just observed the vestiges of zero-temperature phase transition
between $FI$ and $FM$ phases at finite temperature. In general,
there is an extra residual entropy in the interface at zero temperature,
which is responsible for destroying any evidence of zero-temperature
phase transition as soon as temperature increases. However, the phase boundary
between $FI$ and $FM$ has a peculiar property, there is no residual
entropy at the interface, or there is no extra entropy at the boundary \cite{phs-bnd}.
Consequently, the zero-temperature phase transition vestiges survive
at a relatively higher temperature.

\subsection{Magnetization and susceptibility}

So far, we have considered the thermodynamics for null magnetic fields $h_c=h_s=0$.
In this subsection, we investigate the dependence of the free energy with respect
to the fields $h_c$ and $h_s$. For simplicity, we fix $D=-2.480$ and temperatures below, above and at the pseudo-critical temperature $T_p=0.4255221428649$.

\begin{figure}[h!]
    \centering
  \includegraphics[width=7cm]{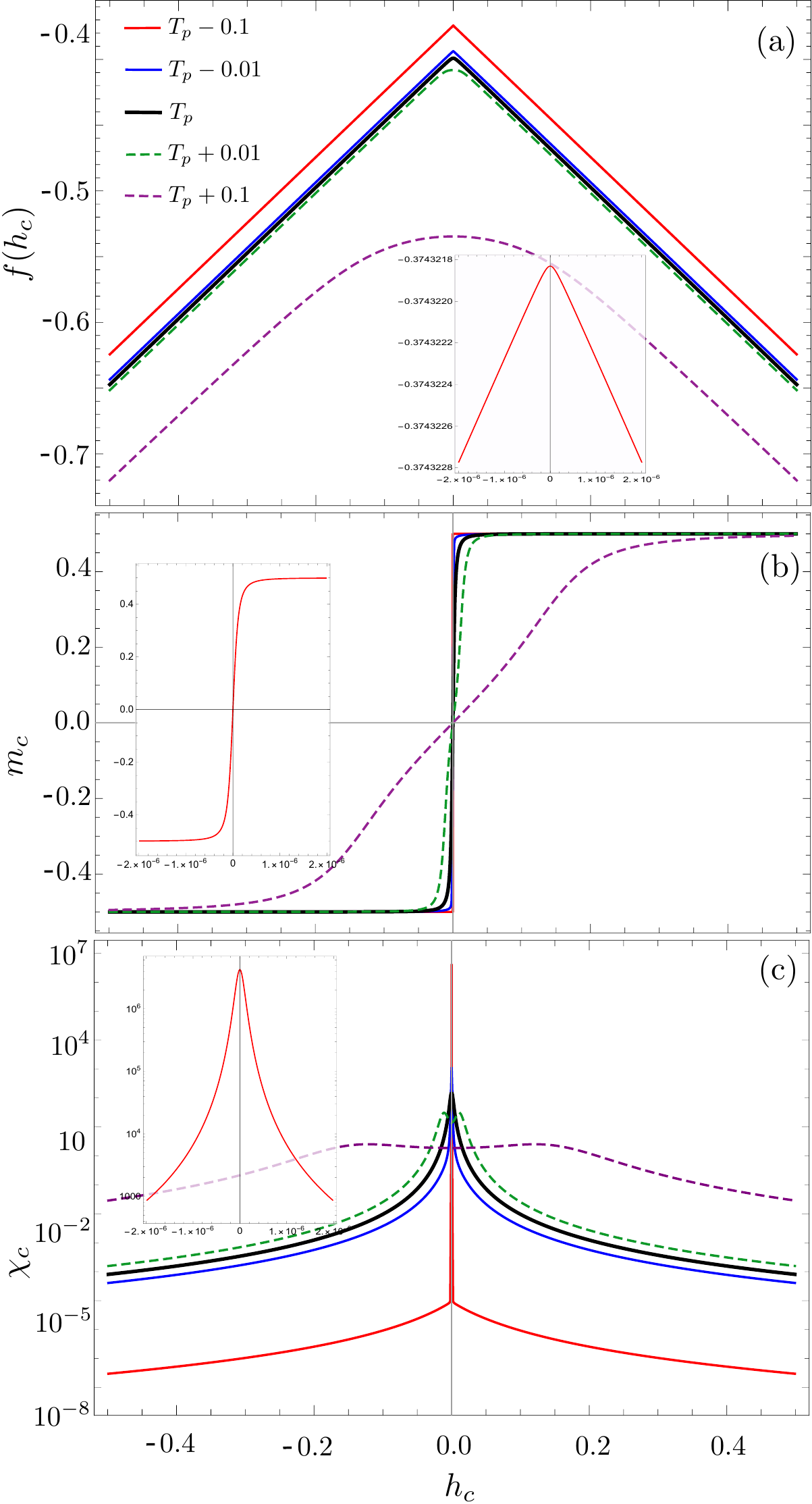}
        \caption{Free energy and its derivatives as functions of $h_c$ for fixed $D=-2.480$. (a) Free energy. (b) Magnetization. (c) Susceptibility.}
        \label{fig:maghc}
\end{figure}

We start by considering $h_s=0$ and varying $h_c$, see fig. \ref{fig:maghc}. The magnetization at given $h_c$ and $T$ can be computed as follows,
\begin{equation}
m_c(h_c)=-\frac{f(T,h_c+\delta h_c)-f(T,h_c-\delta h_c)}{2\delta h_c}\,,
\end{equation}
where $\delta h_c$ is a small step. Here we use $\delta h_c=10^{-20}$ and perform the
numerical calculations using $60$ digits (see appendix \ref{app}).
In panel (a), we plot the dependence of free energy
with respect to the core magnetic field $h_c$. For $T\leq T_p$, we observe an apparent corner at $h_c=0$. This ``corner'', similarly to
our previous discussions, is only apparent, as we highlighted for the extreme case $T=T_p-0.1$. The apparent corner leads to
an abrupt change in the magnetization, as shown in panel (b). In fact, one can see that for temperatures $T<T_p$ (solid line), $T=T_p$ (tick line) and $T>T_p$ (dashed line) the magnetization vanishes smoothly
for zero field $h_c=0$. Therefore, there is no signal of spontaneous magnetization.To recall typical magnetization pictures, see {\it e.g.} \cite{yeo}. Concerning the susceptibility, see panel (c), the strong
changes in the free energy and magnetization reflects into a sharp peak for $\chi_c$. Nevertheless, we can again zoom in very close to $h_c=0$, and realize that in fact the susceptibility is a continuous function of $h_c$, even in the extreme case $T=T_p-0.1$.
It is important to remark that the free energy is a even function of $h_c$ and the null
magnetization at $h_c=0$ is indeed a foreseen result.

\begin{figure}[h!]
    \centering
  \includegraphics[width=8cm]{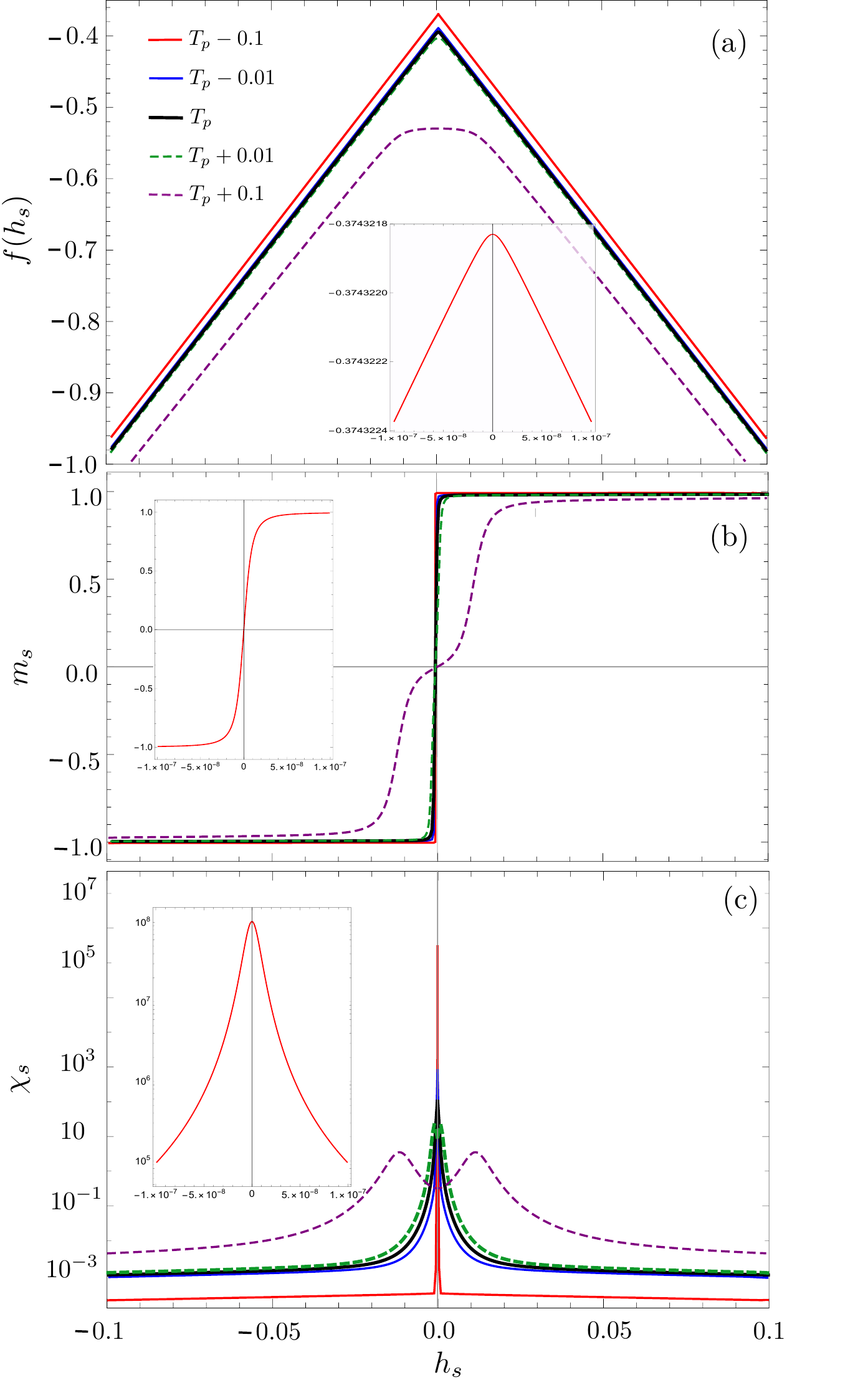}
        \caption{Free energy and its derivatives as functions of $h_s$ for fixed $D=-2.480$. (a) Free energy. (b) Magnetization. (c) Susceptibility.}
        \label{fig:maghs}
\end{figure}

Similar results are obtained by considering $h_c=0$ and varying $h_s$, see fig.\ref{fig:maghs}. 

The fact that approximative methods \cite{Mendes,mendes20} do predict a spontaneous magnetization deserves a deeper investigation from the perspective of
the transfer matrix approach. In order to do that, we turn on a small field $h_c$
, and compute numerically the magnetization around this point using
a very small step $\delta h_c=10^{-20}$, that is,
\begin{equation}
m_c(T)=-\frac{f(T,h_c+\delta h_c)-f(T,h_c-\delta h_c)}{2\delta h_c}\,,
\end{equation}
for different initial fields $h_c$, and similarly the susceptibility. This means that we are slightly far from the ``corner'' in the free energy as a function of $h_c$. Similar calculation is done for $h_s$. 
For these calculations, we choose $D=-2.49$. In order to deal with the small step in the first and second derivatives, we use about 60-80 digits calculations (see appendix \ref{app}). The results are shown in fig.\ref{fig:magvarhsmall}. In panel (a),
we observe that the a small field $h_c=10^{-12}$ is enough to induce a non-zero magnetization in the low-temperature region. Interestingly, for different small fields $h_c$, all curves seems to collapse
as we approach the pesudo-critical temperature. This is confirmed by using a log-scale, see the bottom of panel (a), described by dotted line,
where we observe a change of concavity at $T_p$, although in linear
scale we cannot observe this effect. On the other hand, the magnetic
susceptibility shows an apparent and remarkable peak at $T_p$ for $h_c\sim 10^{-4}-10^{-6}$, inducing us to believe that we are facing a truly phase transition at
$T_p$, but this illusory peak vanishes for $h_c<10^{-6}$, becoming only a substantial 
increase at $T_p$ in the magnetic susceptibility.
For the dashed line, we use the tiny value $h_c=10^{-30}$, with a derivative step of $\delta h_c=10^{-50}$ and perform the calculations using 200 digits. Clearly, the susceptibility leads to a divergence at $T=0$ when $h_c\rightarrow 0$. A similar behavior is observed for the magnetic field $h_s$
of the shell spins, see panel (b).

\begin{figure}[h!]
    \centering
 \includegraphics[width=8cm]{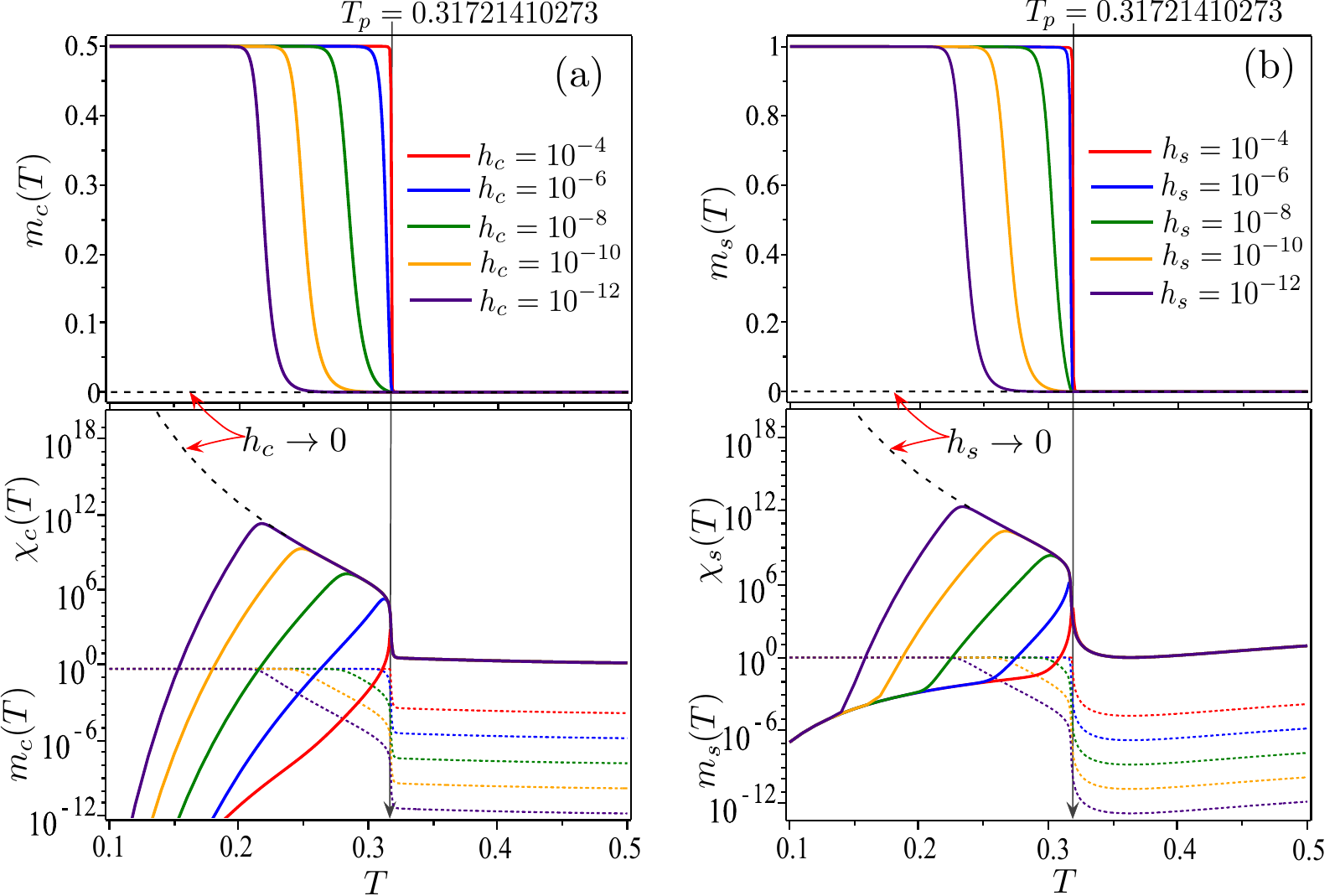}
        \caption{\label{fig:magvarhsmall} Magnetization in linear (top) and log (doted, bottom) scale and susceptibility in log scale for different initial small fields and $\delta h_c=\delta h_s=10^{-20}$. The dashed line in black corresponds to the field $h_c=10^{-30}$. We use here $D=-2.49$.}
\end{figure}

It is worth remarking that the magnetization obtained from the transfer
matrix technique behaves step-like function in temperature. In contrast, the
MFA and MC \cite{Kantar14,Mendes,mendes20} approaches report a rounder function quite
similar to the magnetization of two-dimensional systems.

\subsection{Magnetic quadrupole}

\begin{figure}[h!]
    \centering
         \includegraphics[width=8cm]{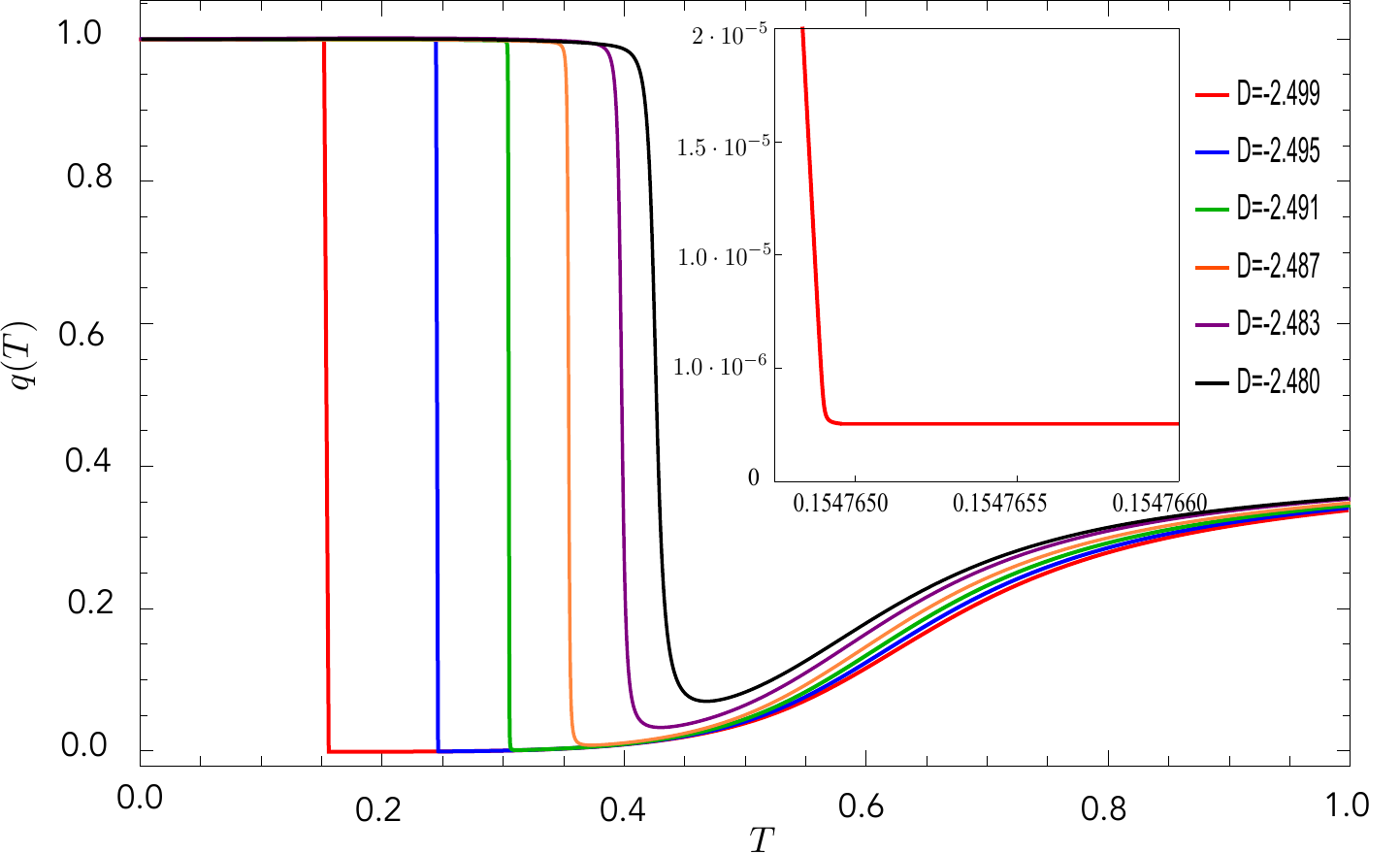}
        \caption{\label{fig:quad}Magnetic quadrupole for various single-ion fields $D$ and $h_c=h_s=0$.}
\end{figure}

We finally investigate the dependence of the magnetic
quadrupole with 
respect to the temperature. The magnetic quadrupole is given
by the derivative of the free energy with respect to the single-ion field $D$
and the results are plotted in fig.\ref{fig:quad}. Similarly to the previous
thermodynamics quantities, we observe a steep change of the function $q(T)$ around the pseudo-critical temperature. As a matter of fact, when we zoom in close to $T_p$ we observe a well behaved function, as enhanced  for $D=-2.499$ in fig.\ref{fig:quad}.
Our results qualitatively look similar to that found in reference
\cite{Mendes}, although the critical temperature does not coincide.

\begin{figure}[h!]
    \centering
         \includegraphics[width=8cm]{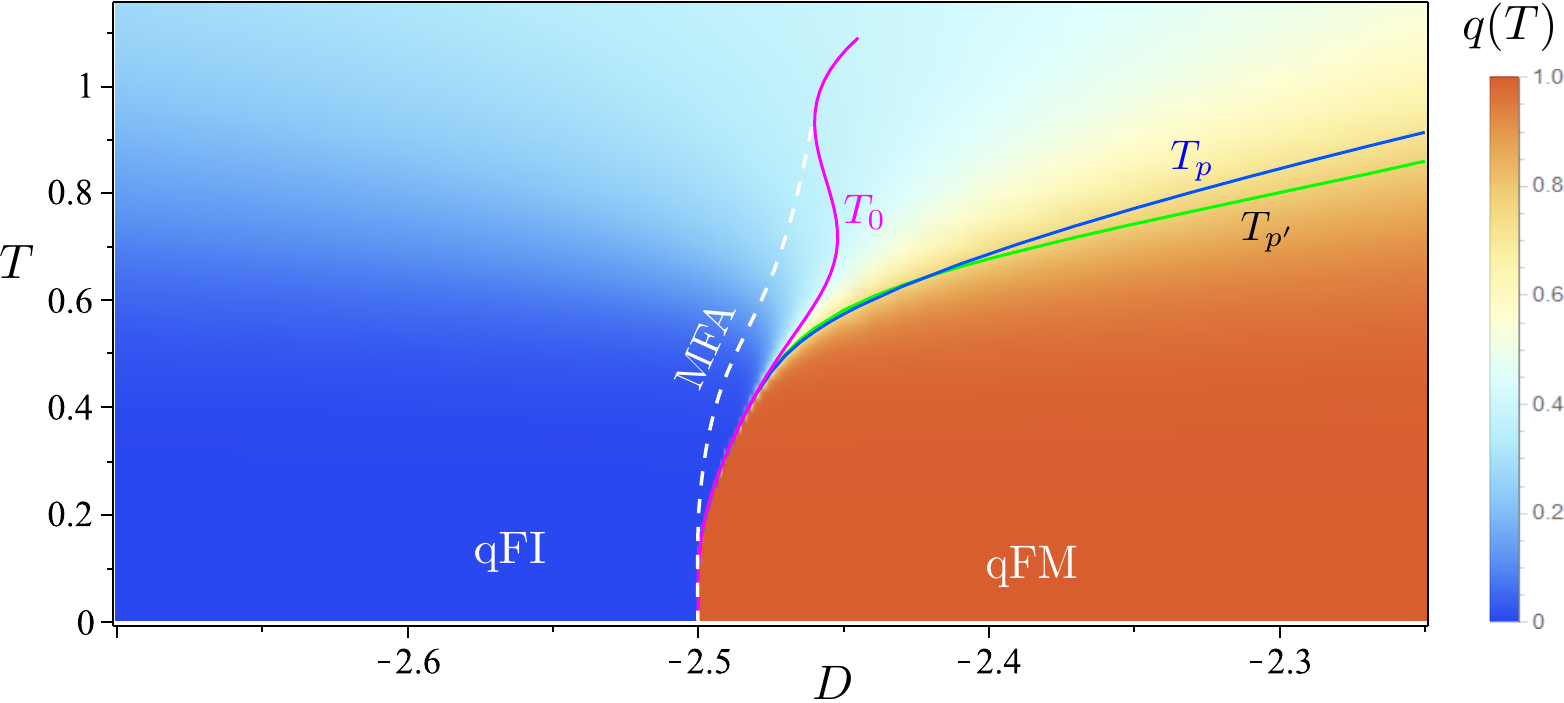}
        \caption{\label{fig:quad2D}Magnetic quadrupole as a function of $T$ and $D$ for $h_c=h_s=0$.}
\end{figure}

An interesting observation is that the magnetic quadrupole as a
function of $T$ and $D$ presents a clear demarcation of the quasi-phases which we discussed previously, see fig.\ref{fig:quad2D}. Definitely, the MFA result \cite{Mendes} deviates from sharp boundary marked by
quasi-phases qFI and qFM.

\section{Conclusion}\label{sec:conclu}

We have analyzed the thermodynamic behavior
of the mixed spin-1/2 and spin-1
hexagonal nanowire system using exact numerical diagonalization
of the transfer matrix. The results indicate the presence of a pseudo-transition, characterized by a steep change of first derivatives of the free energy and by sharp peaks in its second derivatives, around the pseudo-critical temperature, which we have computed for various values
of the crystal field $D$. Nevertheless, by zooming in the thermodynamic quantities around the pseudo-critical temperature we observe that the discontinuities and divergences are only apparent, as expected. This phenomena is linked to the fact the largest eigenvalue of the transfer matrix is quasi degenerate, which means that some of the Boltzmann weights of the transfer matrix are very tiny in the low-temperature region.
In order to observe the non-degeneracy of the eigenvalues, we are required to use high precision numerics. Let us remark that we have considered
a small piece of the model parameters manifold, close to the value $D=-2.5$ and null magnetic fields. It is certainly worth to investigate the thermodynamics beyond this region. The exact numerical diagonalization can be used
to analyze other systems, for instance, a generalized version of the nanowire with spin-$s_1$ core and spin-$s_2$ shell. The limitation is the dimension of the associated transfer matrix.

\section*{Acknowledgments}

RAP was supported by CNPq (grant \# 150829/2020-5). OR and SMS thanks CNPq and FAPEMIG.

\appendix

\section{Numerical derivative significant digits}\label{app}

Performing numerical derivatives require thorough analysis, specially when
we consider the low-temperature region.

In this work we are using centered formula with two points to perform the first order
derivative. To get a reliable result, we take into account the error
of numerical derivatives, see \cite{numd},
\begin{equation}
\left|1-\frac{\bar{f}'(x)}{f'(x)}\right|\lesssim p_{1}\left\{ \frac{h^{2}}{6}+\frac{1}{h}\epsilon\,r_{1}\right\} ,\label{eq:1drv-cnd}
\end{equation}
here $f'(x)$ is the first order derivative of $f(x),$ while 
\begin{equation}
\bar{f}'(x)\equiv\frac{f(x+h)-f(x-h)}{2h},
\end{equation}
is the numeric derivative and
\begin{equation}
p_{1}=\frac{|f'''(x)|}{|f'(x)|},\quad r_{1}=\frac{|f(x)|}{|f'''(x)|},
\end{equation}
and $\epsilon$ is the machine epsilon.

Basically, the root of this error lies in the truncation error proportional
to $h^{2}$ and the round-off error proportional to $h^{-1}$; this
means that we cannot carry out $h\rightarrow0$ as small as we want.
Therefore, by minimizing the error the optimal value for $h$ leads
to the following relation 
\begin{equation}
\epsilon=\frac{h^{*3}}{3r_{1}}\,.
\end{equation}
At optimal $h^{*}$ the relation \eqref{eq:1drv-cnd} becomes
\begin{equation}
\left|1-\frac{\bar{f}'(x)}{f'(x)}\right|\approx p_{1}\left\{ \frac{h^{*2}}{6}+\frac{1}{h^{*}}\epsilon\,r_{1}\right\} =\frac{p_{1}}{2}h^{*2}.\label{eq:at opt-h}
\end{equation}
The significant digits of numerical derivative is defined by
\begin{alignat*}{1}
n_{o}= & -\log_{10}\Bigl|1-\frac{\bar{f}'(x)}{f'(x)}\Bigr|\,.
\end{alignat*}
We can express $n_{h}=-\log_{10}(h^{*}),$ as a function of $n_{\epsilon}$
($\epsilon=10^{-n_{\epsilon}}$), obtaining,
\begin{equation}
n_{h}=\frac{1}{3}n_{\epsilon}-\frac{1}{3}\log_{10}(3r_{1}).
\end{equation}
Typically $r_{1}$ should be of the order $r_{1}\sim1$, but when
$r_{1}\gg1$, special attention needs to be taken when performing
numerical derivatives with a double-precision number. This means that
for a double-precision number, we have $n_{\epsilon}\sim15$, so the
optimal value of $h^{*}$ should not be lower than $h^{*}\sim10^{-5}$.

Rewriting eq.\eqref{eq:at opt-h}, we have 
\begin{equation}
n_{o}\approx2n_{h}-\log_{10}(\tfrac{p_{1}}{2}).
\end{equation}

As a consequence, the accuracy of the numerical differentiation
could have no more than $n_{o}\lesssim2n_{\epsilon}/3\sim10$ significant
digits. Of course, one could improve this result by using more sophisticated numerical
derivative, like centered formula with 4 or more points, but this is time consuming, and one would not go far than double precision number accuracy. Usually, $p_{1}$ is of order
$p_{1}\sim 1$, but in low-temperature regions, $p_{1}$ should be
$p_{1}\gg1$. Accuracy might be even worse, making it difficult to
obtain a precise result. The only possibility to improve the numerical
derivative considerably is by increasing the significant decimal digits.
For example, assuming $n_{\epsilon}\approx30$, the corresponding
optimal $h^{*}$ should be of order $h^{*}\approx10^{-10}$ with numerical
precision of order $n_{o}\sim15$ although this precision might be significantly
lower than 15 digits for lower temperature. 

A similar analysis may be done for the second order derivative. Here we use central formula with 3 points,
\begin{equation}
\left|1-\frac{\bar{f}''(x)}{f''(x)}\right|\lesssim p_{2}\left\{ \frac{h^{2}}{12}+\frac{1}{h^{2}}\epsilon\,r_{2}\right\} ,\label{eq:df2}
\end{equation}
here $f''(x)$ is the second order derivative of $f(x),$ while 
\begin{equation}
\bar{f}''(x)\equiv\frac{f(x+h)-2f(x)+f(x-h)}{h^{2}},
\end{equation}
is the numeric derivative with 
\begin{equation}
p_{2}=\frac{|f^{(iv)}(x)|}{|f'(x)|},\quad r_{2}=\frac{|f(x)|}{|f^{(iv)}(x)|}\,.
\end{equation}
As we can see in \eqref{eq:df2}, $h$ cannot be set arbitrarily small 
because of numerical precision $\epsilon$. Instead, minimizing the error
of \eqref{eq:df2} we have the optimal value,
\begin{equation}
\epsilon\approx\frac{h^{*4}}{36r_{2}}\,.
\end{equation}
For the optimal $h$, the eq.\eqref{eq:df2} results in
\begin{equation}
\left|1-\frac{\bar{f}''(x)}{f''(x)}\right|\approx p_{2}\left\{ \frac{h^{2}}{12}+\frac{1}{h^{2}}\epsilon\,r_{2}\right\} =\frac{p_{2}}{9}h^{*2}\,.
\end{equation}

\vspace{0.1cm}

\begin{figure}[h!]
\includegraphics[scale=0.5]{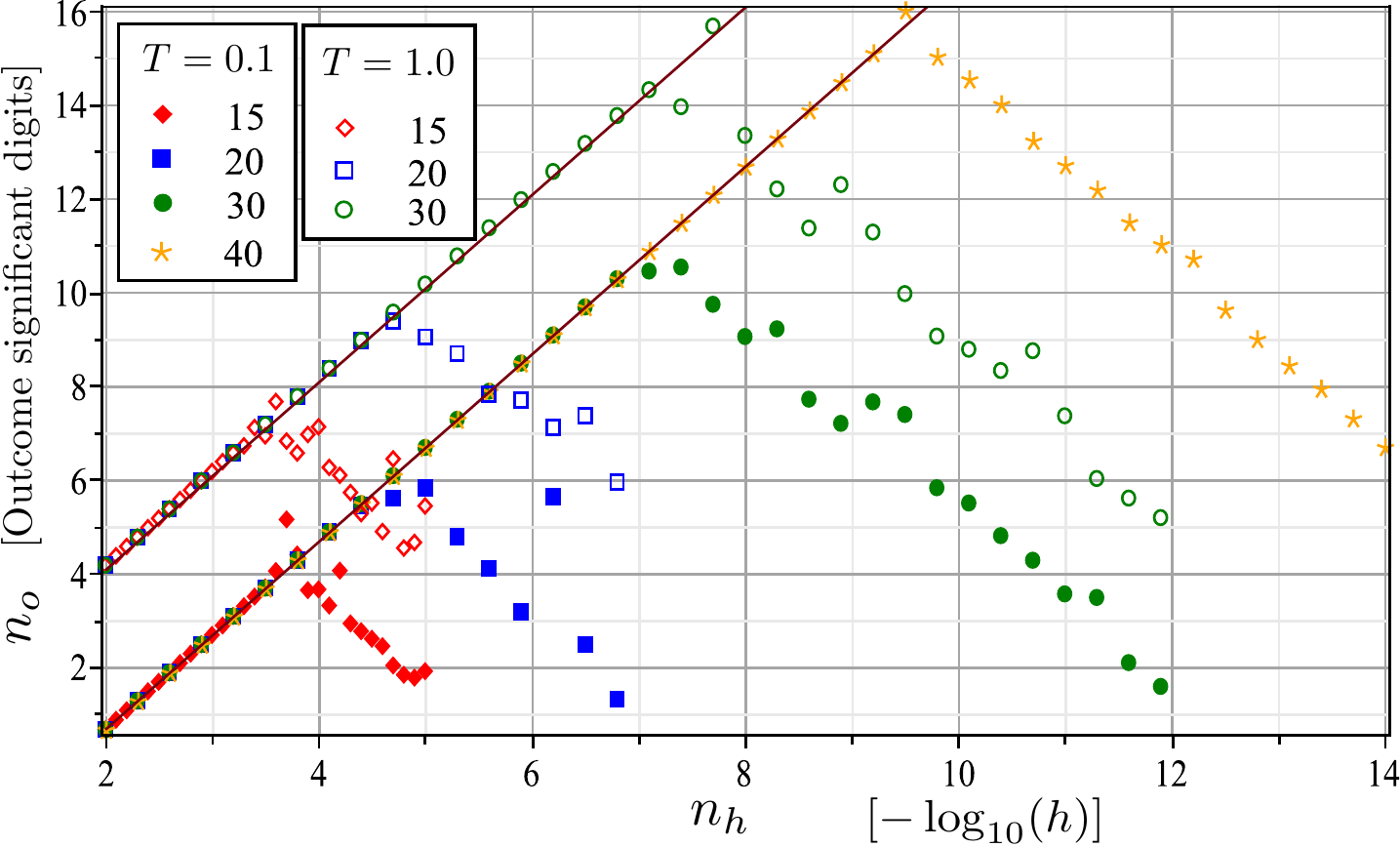}\caption{\label{fig:drv-err}Significant digits $n_{o}$ of specific heat as
function of $n_{h}$, for $D=-2.499$ and null magnetic field. Upper
solid line {[}$n_{o}\approx2n_{h}+0.1${]} corresponds to $T=1$,
and lower solid line {[}$n_{o}\approx2n_{h}-3.3${]} is given for
$T=0.1$. }
\end{figure}

Therefore, the precision of significant digits becomes
\begin{equation}
n_{o}\approx2n_{h}-\log_{10}(\frac{p_{2}}{9}).
\end{equation}

In order to illustrate the the significant digits $n_{o}$ as a function
of $n_{h}$, we show in fig.\ref{fig:drv-err} the accuracy
of the specific heat assuming $D=-2.499$ and null magnetic fields. The
upper solid line is given by $n_{o}\approx2n_{h}+0.1$, which corresponds
to $T=1$. Here $p_{2}$ can be obtained from fig. \ref{fig:drv-err}, which
is of order $p_{2}\sim7.1$. In this case the double precision number
$n_{\epsilon}\approx15$ leads to an optimal significant digits $n_{o}\sim7$
of specific heat, and by extending to $n_{\epsilon}\sim20$ we get
$n_{o}\sim9$ significant digits, while for $n_{\epsilon}\sim30$
we get $n_{o}\sim14$. However, for the lower temperature $T=0.1$, the
solid line is given by $n_{o}\approx2n_{h}-3.3$, and the corresponding
$p_{2}$ becomes $p_{2}\sim1.8\times10^{4}$. This factor influences
significantly in the specific heat precision falling roughly in 3
significant digits. For lower temperatures the accuracy would decrease
even more dramatically.

\end{document}